\documentclass[article,nojss]{jss}\usepackage[]{graphicx}\usepackage[]{xcolor}
\makeatletter
\def\maxwidth{ %
  \ifdim\Gin@nat@width>\linewidth
    \linewidth
  \else
    \Gin@nat@width
  \fi
}
\makeatother

\definecolor{fgcolor}{rgb}{0.345, 0.345, 0.345}

\usepackage{framed}
\makeatletter
 {\par\unskip\endMakeFramed%
 \at@end@of@kframe}
\makeatother

\definecolor{shadecolor}{rgb}{.97, .97, .97}
\definecolor{messagecolor}{rgb}{0, 0, 0}
\definecolor{warningcolor}{rgb}{1, 0, 1}
\definecolor{errorcolor}{rgb}{1, 0, 0}
\usepackage{alltt}
\usepackage{orcidlink,thumbpdf,lmodern}
\usepackage{framed}
\usepackage{latexsym}
\usepackage{amssymb}
\usepackage{amsmath}
\usepackage{amsfonts}
\usepackage{mathtools}
\usepackage{bm, bbm}
\usepackage{algorithm, algpseudocode}
\usepackage{rotating}
\usepackage{multirow}
\usepackage{booktabs}
\usepackage{float}

\newcommand{\indep}{\perp \!\!\! \perp}
\newcommand{\der}[2]{\frac{d {#1}} {d{#2}}}

\newcommand{\red}[1]{\textcolor{black}{#1}}

\newcommand{\fct}[1]{\code{#1()}}

\author{Thomas A. Trikalinos~\orcidlink{0000-0002-3990-1848}\\Brown University
   \And Yuliia Sereda~\orcidlink{https://orcid.org/0000-0002-4017-4561}\\Brown University}
\Plainauthor{Thomas A. Trikalinos, Yuliia Sereda}

\title{\pkg{nhppp}: Simulating Nonhomogeneous Poisson Point Processes in \proglang{R}}
\Plaintitle{nhppp: Simulating Nonhomogeneous Poisson Point Processes in R}
\Shorttitle{Simulating NHPPPs in R}

\Abstract{
  We introduce the \pkg{nhppp} package for simulating events from one-dimensional non-homogeneous Poisson point processes (NHPPPs) in \proglang{R} \red{fast and with a small memory footprint. We developed it to facilitate the sampling of event times in discrete event and statistical simulations. The package's} functions are based on three algorithms that provably sample from a target NHPPP: the time-transformation of a homogeneous Poisson process (of intensity one) via the inverse of the integrated intensity function; the generation of a Poisson number of order statistics from a fixed density function; and the thinning of a majorizing NHPPP via an acceptance-rejection scheme. We present a study of numerical accuracy and time performance of the algorithms. \red{We illustrate use} with simple reproducible examples.
}

\Plainkeywords{stochastic point processes, counting processes, discrete event simulation, time-to-event simulation, R}

\Address{
  TA Trikalinos\\
  Department of Health Services, Policy \& Practice\\
  \emph{and}\\
  Department of Biostatistics\\
  School of Public Health\\
  Brown University\\
  RI 02912, USA\\
  E-mail: \email{thomas\_trikalinos@brown.edu}
}
\IfFileExists{upquote.sty}{\usepackage{upquote}}{}
\begin{document}

\section{Introduction} \label{sec:intro}

It is often desirable to simulate series of events (stochastic point processes) so that the intensity of their occurrence varies over time. Examples include events such as the occurrence of death and occurrences of symptoms, infections, or tumors over a person's lifetime. The non-homogeneous Poisson point process (NHPPP), which generalizes the simpler homogeneous-Poisson, Weibull, and Gompertz point processes, is a widely used model for such series of events. NHPPPs can model complicated event patterns given a suitable intensity function. They are, therefore, useful in statistical and mathematical model simulation.

An NHPPP has the properties that the number of events in all non-overlapping time intervals are independent random variables and that, within each time interval, the number of events is Poisson distributed. Thus an NHPPP is a memoryless point process. A large number of phenomena may reasonably conform with these properties.%

The \pkg{nhppp} package in \proglang{R} contains functions for the simulation of NHPPPs over a one-dimensional carrier space, which we will take to represent time~\citep{nhppp-package, trikalinos2024nhppp}. \red{Table~\ref{tab:nhppp_functions} summarizes the functions implemented in \pkg{nhppp} as of version 0.1.4.} You can install the development version of \pkg{nhppp} with
\begin{Schunk}
\begin{Sinput}
R> # install.packages("devtools")
R> devtools::install_github("bladder-ca/nhppp")
\end{Sinput}
\end{Schunk}
or the release version with
\begin{Schunk}
\begin{Sinput}
R> install.packages("nhppp")
\end{Sinput}
\end{Schunk}

% Please add the following required packages to your document preamble:
% \usepackage{booktabs}
% \usepackage{multirow}
% \usepackage{graphicx}

\begin{table}[ht!]
\caption{\textbf{Functions in \pkg{nhppp}}}
\label{tab:nhppp_functions}
\vspace{5pt}
\resizebox{\textwidth}{!}{%
\begin{tabular}{@{}lll@{}}
\toprule
\textbf{Intensity function}
& \textbf{Function name}
& \textbf{Function simulates...}
\\ \midrule
\multirow{4}{*}{Constant}
& \fct{ppp\_n}
& exactly $n$ events in $[a, b)$
\\
& \fct{ppp\_next\_n} 
& the next $n$ events in $[a, \infty)$                                   \\
& \fct{ppp\_sequential}, \fct{ppp\_orderstat}  
& $\ge 0$ events in $[a,b)$
\\
& \fct{ztppp} 
& $\ge 1$ events in $[a, b)$
\\
\addlinespace
\multirow{6}{*}{\begin{tabular}[c]{@{}l@{}}Time-varying, \\ special cases, \\ non-vectorized\end{tabular}} 
& \fct{draw\_sc\_linear} 
& $\ge 0$ events in $[a,b)$ from $\lambda(t) = \alpha + \beta  t$
\\
& \fct{draw\_sc\_loglinear} 
& $\ge 0$ events in $[a,b)$ from $\lambda(t) = \exp(\alpha + \beta t)$   
\\
& \fct{draw\_sc\_step} 
& \begin{tabular}[t]{@{}l@{}} $\ge 0$ events in $[a, b)$ from piecewise constant \\$\lambda(t)$ with uneven intervals\end{tabular} 
\\
& \fct{draw\_sc\_step\_regular}
& \begin{tabular}[t]{@{}l@{}} $\ge 0$ events in $[a, b)$ from piecewise constant \\$\lambda(t)$ with regular intervals\end{tabular}
\\
& \fct{ztdraw\_sc\_linear}
& $\ge 1$ events in $[a,b)$ from $\lambda(t) = \alpha + \beta  t$
\\
& \fct{ztdraw\_sc\_loglinear}
& $\ge 1$ events in $[a,b)$ from $\lambda(t) = \exp(\alpha + \beta  t)$ 
\\
\addlinespace
\multirow{2}{*}{\begin{tabular}[c]{@{}l@{}}Time varying, \\ special cases,  \\ vectorized\end{tabular}}
& \fct{vdraw\_sc\_step\_regular}
& \begin{tabular}[t]{@{}l@{}} $\ge 0$ events in $[a, b)$ from piecewise constant \\$\lambda(t)$ with regular intervals\end{tabular}
\\
& \fct{vztdraw\_sc\_step\_regular}
& \begin{tabular}[t]{@{}l@{}} $\ge 1$ events in $[a, b)$ from piecewise constant \\$\lambda(t)$ with regular intervals\end{tabular}
\\
\addlinespace
\multirow{8}{*}{\begin{tabular}[c]{@{}l@{}}Time-varying,  \\ general case,\\ non-vectorized\end{tabular}}      

& \fct{draw}
& (wrapper function)                                                     
\\
& \fct{draw\_cumulative\_intensity\_inversion}  
& $\ge 0$ events in $[a,b)$ from $\Lambda(t), \Lambda^-1(t)$
\\
& \fct{draw\_cumulative\_intensity\_orderstats} 
& $\ge 0$ events in $[a,b)$ from $\Lambda(t), \Lambda^-1(t)$
\\
& \fct{draw\_intensity} 
& $\ge 0$ events in $[a,b)$ from $\lambda(t), \lambda_*(t)$
\\
& \fct{draw\_intensity\_step}
& \begin{tabular}[t]{@{}l@{}} $\ge 0$ events in $[a,b)$ from $\lambda(t)$ and\\ piecewise constant $\lambda_*(t)$ \end{tabular}
\\
& \fct{ztdraw\_cumulative\_intensity}           
& $\ge 1$ events in $[a,b)$ from $\Lambda(t), \Lambda^-1(t)$
\\
& \fct{ztdraw\_intensity} 
& $\ge 1$ events in $[a,b)$ from $\lambda(t), \lambda_*(t)$
\\
& \fct{ztdraw\_intensity\_step} 
& \begin{tabular}[t]{@{}l@{}} $\ge 1$ events in $[a,b)$ from $\lambda(t)$ and\\ piecewise constant $\lambda_*(t)$ \end{tabular} 
\\
\addlinespace
\multirow{3}{*}{\begin{tabular}[c]{@{}l@{}}Time varying,\\ general case, \\ vectorized\end{tabular}}
& \fct{vdraw}
& (wrapper function)
\\
& \fct{vdraw\_intensity\_step\_regular}
& \begin{tabular}[t]{@{}l@{}} $\ge 0$ events in $[a,b)$ from $\lambda(t)$ and\\ piecewise constant $\lambda_*(t)$ \end{tabular} 
\\
& \fct{vztdraw\_intensity\_step\_regular}
& \begin{tabular}[t]{@{}l@{}} $\ge 1$ events in $[a,b)$ from $\lambda(t)$ and\\ piecewise constant $\lambda_*(t)$ \end{tabular} 
\\
\addlinespace
(Helper function)
& \fct{get\_step\_majorizer}
& (obtains piecewise constant $\lambda_*(t)$ from $\lambda(t)$)                                             
\\ \bottomrule
\end{tabular}%
} %resizebox
\begin{flushleft}
The table pertains to version 0.1.4 of \pkg{nhppp}. $\lambda(t)$ is an intensity function, 
$\lambda_*(t)$ a majorizer function for $\lambda(t)$, $\Lambda(t)$ the integrated intensity function, and $\Lambda^{-1}(t)$ the inverse function (preimage) of $\Lambda(t)$.
\end{flushleft}
\end{table}

We review NHPPPs in Section~\ref{sec:review} and algorithms for sampling from constant rate Poisson point processes in Section~\ref{sec:sample-ppp}. We introduce the three sampling algorithms that are implemented in the package in Section~\ref{sec:general-sampling}. We discuss special functional forms for the intensity function (constant, piecewise constant, linear, and log-linear) in Section~\ref{sec:special_cases}. We describe \pkg{nhppp} versus other \proglang{R} packages that can simulate from one-dimensional NHPPPs in Section~\ref{sec:other-R-packages} and present a numerical study in Section~\ref{sec:illustrations}. We summarize in Section~\ref{sec:summary}.

\section{The Poisson point process} \label{sec:review}
\subsection{Definition}
The Poisson point process is a stochastic series of events on the real line. For some sequence of events, let $N(t, t + \Delta t)$ be the number of events in the interval $(t, t  + \Delta t]$. If for some positive intensity $\lambda$ and, as ${\Delta t \rightarrow 0}$,
\begin{equation}\label{eq:definition}
    \begin{aligned}
    \Pr[N(t, t + \Delta t) = 0] =&\  1 - \lambda \Delta t +  o(\Delta t), \\
    \Pr[N(t, t + \Delta t) = 1] =&\  \lambda \Delta t +  o(\Delta t), \\
    \Pr[N(t, t + \Delta t) >1] =&\  o(\Delta t),\text{ and } \\
    N(t, t + \Delta t) \indep&\ N(0, t),
    \end{aligned}
\end{equation}
then that sequence of events is a Poisson point process. In Equation~\eqref{eq:definition}, the third statement demands that events occur one at a time. The fourth statement implies that the process is memoryless: For any time $t_0$, the behavior of the process is independent to what happened before that time.

\subsection{Homogeneous Poisson point process and counting process}\label{sec:ppp-intro}
Assume that the next event after time $t_0$ happens at time $t_0 + X$. It follows from the above definition \citep[par. 4.1]{cox1965theory} that, for a constant $\lambda$, $X$ is exponentially distributed
\begin{equation}\label{eq:X_PPP}
X \sim \text{Exponential}(\lambda),
\end{equation}
and that the number of events is Poisson distributed over the compact interval $(a, b]$, i.e.,
\begin{equation}\label{eq:N_PPP}
N(a, b) \sim \text{Poisson}(\lambda (b-a)).
\end{equation}

Equation~\eqref{eq:X_PPP} generates the homogeneous Poisson point process ${Z_1 = t_0 + X_1, Z_2 = Z_1 + X_2, \dots}$, where $Z_i$ is the time of arrival of event $i$ and $X_i$ the inter-arrival times. We will use $Z_{(j)}$ to denote the event in position $j$ when events are ordered in increasing time.
Equation \eqref{eq:N_PPP} describes the corresponding (dual) counting process
${N_1 = N(t_0, Z_1)}, {N_2 = N(t_0, Z_2), \dots}$, where $N_i$ is the total number of events from time $t_0$ to time $Z_i$. The point process (the sequence $[Z_i]$ of event times) and the counting process (the sequence $[N_i]$ of cumulants) are two sides of the same coin.

Sampling from the constant rate point process in~\eqref{eq:X_PPP} is discussed in Section~\ref{sec:sample-ppp}.

\subsection{Non homogeneous Poisson point process and counting process}\label{sec:nhppp-intro}
When the intensity function changes over time, the homogeneous Poisson point process generalizes to its non-stationary counterpart, an NHPPP, with intensity function $\lambda(t) > 0$. For details see \citet[par 4.2]{cox1965theory}. Then, the number of events over the interval $(a, b]$ becomes
\begin{equation}\label{eq:N_NHPPP}
N(a, b) \sim \text{Poisson}(\Lambda(a, b)),
\end{equation}
where $\Lambda(a, b) = \int_a^b \lambda(t) \ dt$ is the integrated intensity or cumulative intensity of the NHPPP. Equation~\eqref{eq:N_NHPPP} describes the counting process of the NHPPP, which in turn implies a stochastic point process -- a distribution of events over time.

Here the simulation task is to sample event times from the point process that corresponds to intensity function $\lambda(t)$, or equivalently, to the integrated intensity function $\Lambda(t) = \int_0^t \lambda(s) \ ds$ (Section~\ref{sec:general-sampling}).
(With some abuse of notation, we define $\Lambda(t) \coloneqq \Lambda(0, t)$ when  $a=0$.)

\subsubsection{A note on zero intensity processes}
In~\eqref{eq:definition}, $\lambda$ is strictly positive but in~\pkg{nhppp} we allow it to be non-negative. If $\lambda = 0$, ${\Pr[N(t, t + \Delta t) = 0] = 1}$ and ${\Pr[N(t, t + \Delta t) \ge 1] = 0}$. This means that no events occur and the stochastic point process in the interval $(t, t + \Delta t]$ is denegerate. Allowing $\lambda(t) \ge 0$ has no bearing on the results of simulations. If
\begin{equation*}
    \lambda(t)  \begin{cases}
    >0, \text{ for } t \in (a, b] \\
    =0, \text{ for } t \in (b, c] \\
    >0, \text{ for } t \in (c, d]
    \end{cases}
\end{equation*}
we can always ignore the middle interval in which no events happen.

\subsection{Properties that are important for simulation}\label{sec:properties}
\subsubsection{Composability and decomposability of NHPPPs}
The definition~\eqref{eq:definition} implies that NHPPPs are composable~\cite[par. 4.2]{cox1965theory}: merging two NHPPPs with intensity functions $\lambda_1(t), \lambda_2(t)$ yields a new NHPPP with intensity function $\lambda(t) = \lambda_1(t) + \lambda_2(t)$. The reciprocal is also true: one can decompose an NHPPP with intensity function $\lambda(t)$ into two NHPPPs, one with intensity function $\lambda_1(t) < \lambda(t)$ and one with intensity function ${\lambda_2(t) = \lambda(t)-\lambda_1(t)}$. An induction argument extends the above to merging and decomposing three or more processes.

The composability and decomposability properties are important for simulation because they
\begin{itemize}
    \item give the flexibility to simulate several parallel NHPPPs independently versus to merge them, simulate from the merged process, and then attribute the realized events to the component processes by assigning the $i$-th event to the $j$-th process with probability $\lambda_j(Z_i) / \lambda(Z_i)$, where $\lambda(t) = \sum \lambda_j(t)$.
    \item motivate a general sampling algorithm (Algorithm~\ref{alg:NHPPP_thinning}, ``thinning'' in~\cite{lewis1979thinning}) that simulates a target NHPPP with intensity $\lambda_1(t)$ by first drawing events from an easy-to-sample NHPPP with intensity $\lambda(t) >\lambda_1(t)$, and then accepts sample $i$ with probability $\lambda_1(Z_i)/\lambda(Z_i)$.
\end{itemize}

\subsubsection{Transformations of the time axis}
Strictly monotonic transformations of the carrier space of an NHPPP yield an NHPPP~\citep{Cinlar1975inversion}. Consider an NHPPP with intensity functions $\lambda(t)$ and a strictly monotonic transformation of the time axis $u: t \mapsto \tau$ that is differentiable once almost everywhere. On the transformed time axis the point process is an NHPPP with intensity function
\begin{equation}\label{eq:transform}
    \rho(\tau) = \lambda(\tau) \left ( \der{u}{t} \right )^{-1}.
\end{equation}

This property is important for simulation because
\begin{itemize}
    \item it motivates the use of another general sampling algorithm (Algorithm~\ref{alg:NHPPP_inversion}, ``time transformation'' or ``inversion'', see~\cite{Cinlar1975inversion}): A smart choice for $u$ yields an easy to sample point process. The event times in the original time scale can be obtained as $Z_i = u^{-1}(\zeta_i)$, where $\zeta_i$ is the $i$-th event in the transformed time axis and $u^{-1}$ is the inverse function of $u$.
    \item given that at least $i$ events have realized in the time interval $(a, b]$, it makes it possible to draw events ${Z_{(j)}, j<i}$ given event $Z_{(i)}$. This is useful for simulating earlier events conditional on the occurrence of a subsequent event. Choosing $u(t) := Z_{(i)} - t$ makes the time count backwards from $Z_{(i)}$. In this reversed clock we draw as if in forward time exactly $i-1$ events $\zeta_{(1)}, \zeta_{(2)}, \dots, \zeta_{(i-1)}$. Back transforming yields all preceding events.
\end{itemize}

Table~\ref{tab:simtasks} summarizes the common simulation tasks, such as simulating single events (at most one, exactly one), a series of events (possibly demanding the occurrence of at least one event), or the occurrence of a prior (event $i-1$ given $Z_{(i)}$). The \pkg{nhppp} package implements functions to simulate these tasks for general $\lambda(t)$ or $\Lambda(t)$.

\begin{table}[ht!]
    \caption{\textbf{Common simulation needs in discrete event simulation}}\label{tab:simtasks} 
    \vspace{5pt}
    \centering
    \begin{tabular}{lp{2cm}lp{1.8cm}p{6cm}}
    \toprule
    \textbf{\#} & \textbf{Sampling task} & \textbf{Sampled times} & \textbf{Number of sampled events} & \textbf{Example}  \\ 
    \midrule
    I & Any next event&  $\{\}$  or $\{Z_{(1)}\}$&  0 or 1& Single event that may (or may not) occur in the interval: death, progression from Stage I to Stage II cancer. \\ 
    \addlinespace
    II & Exactly one next event&  $\{Z_{(1)} \}$&  1& Single event which must occur in the interval: death from any cause in a lifetime-horizon simulation. \\ 
    \addlinespace
    III & Any and all events&  $\{\}$ or $\{Z_{(1)}, Z_{(2)}, \dots \}$&  $\ge 0$& Zero, one, or more events: emergence of one or more bladder tumors. \\ 
    \addlinespace
    IV & At least one next event&  $\{Z_{(1)} , Z_{(2)}, \dots \}$&  $\ge 1$& One or more events: emergence of bladder tumors when simulating only patients with bladder tumors. \\ 
    \addlinespace
    V & Event $i-1$ given $Z_{(i)}$ & $\{Z_{(i-1)}\}$ & 1 & Find the previous event when simulating conditional on a future event: time of symptom onset given the time of symptom-driven diagnosis; onset of Stage I cancer given progression from Stage I to Stage II cancer. \\ 
    \bottomrule
    \end{tabular}
    \begin{flushleft}
    All listed tasks involve sampling events over the interval $(a, b]$ with known $\lambda(t)$ or $\Lambda(t)$.
    \end{flushleft}
\end{table}

\section{Sampling the constant rate Poisson process}\label{sec:sample-ppp}

Sampling the constant rate Poisson process is straightforward. Algorithms~\ref{alg:PPP_t} and~\ref{alg:PPP_order_stats} are two ways to sample event times in interval $(a, b]$ with constant intensity $\lambda$. Algorithm~\ref{alg:PPP_conditional} describes sampling event times conditional on observing at least $k$ events within the interval of interest.

\subsection{Sequential sampling}\label{sec:PPP_t}

Algorithm~\ref{alg:PPP_t} samples events sequentially, using the fact that the inter-event times $X_i$ are exponentially distributed with mean $\lambda^{-1}$~\cite[par. 4.1]{cox1965theory}. It involves generation only of exponential random variates, which is cheap on modern hardware. To sample at most $k$ events, change the condition for the while loop in line 3 to
\begin{center}
\textbf{while} {$t <b  \ \& \  |\mathcal{Z}| < k$} \textbf{do}.
\end{center}

\begin{algorithm}[ht!]
\caption{Sequential sampling of events in interval $(a, b]$ with constant intensity $\lambda$. }\label{alg:PPP_t}
\begin{algorithmic}[1]
\Require $t \in (a, b]$
%\Ensure $b > a$
\State $t \gets a$
\State $\mathcal{Z} \gets \emptyset$ \Comment{$\mathcal{Z}$ is an ordered set}
\While{$t < b$}   \Comment{Up to $k$ earliest points: \textbf{while} {$t <b  \ \& \  |\mathcal{Z}| < k$} \textbf{do}}
    \State $X \gets X \sim \textrm{Exponential}(\lambda^{-1})$ \Comment{Mean-parameterized}
    \State $t \gets t + X$
    \If{$t < b$}
        \State $\mathcal{Z} \gets \mathcal{Z} \cup \{t\} $
    \EndIf
\EndWhile
\State
\Return {$\mathcal{Z}$}
\end{algorithmic}
\end{algorithm}

The package's \fct{ppp\_sequential} function implements constant-rate sequential sampling that returns a vector with zero or more event times in the interval $[a, b)$. The \code{range\_t} argument is a two-values vector with the bounds $a, b$. 
Setting the optional argument \code{atmost1} to \code{TRUE} from its default value of \code{FALSE} returns the first event or an empty vector, depending on whether at least one event is drawn in the interval.

\begin{Schunk}
\begin{Sinput}
R> library("nhppp")
R> ppp_sequential(range_t = c(7, 10), rate = 1, atmost1 = FALSE)
\end{Sinput}
\begin{Soutput}
[1] 7.673885 8.650502 9.011229 9.407575
\end{Soutput}
\end{Schunk}

Most \pkg{nhppp} functions can accept a user provided random number stream object.

\begin{Schunk}
\begin{Sinput}
R> library("rstream")
R> S <- new("rstream.mrg32k3a")
R> ppp_sequential(range_t = c(7, 10), rate = 1, rng_stream = S)
\end{Sinput}
\begin{Soutput}
[1] 8.793702
\end{Soutput}
\end{Schunk}

\subsection{Sampling using order statistics}\label{sec:PPP_order_stats}
\begin{algorithm}[h!]
\caption{Sampling events in interval $(a, b]$ with constant intensity $\lambda$ using order statistics. }\label{alg:PPP_order_stats}
\begin{algorithmic}[1]
\Require $t \in (a, b]$
%\Ensure $b > a$
\State $N \gets N \sim \textrm{Poisson}\big(\lambda (b-a)\big)$
\State $t \gets a$
\State $\mathcal{Z} \gets \emptyset$ \Comment{$\mathcal{Z}$ is an ordered set}
\If{N > 0}
    \For{$i \in [N]$}:
        \State $U_i \gets U_i \sim \textrm{Uniform(0, 1)}$ \Comment{Generate order statistics}
        \State $\mathcal{Z} \gets \mathcal{Z} \cup \{a + (b-a) U_i\} $ 
    \EndFor
    \State $\mathcal{Z} \gets \textrm{sort}(\mathcal{Z})$ 
\EndIf
\State
\Return{$\mathcal{Z}$} \Comment{Up to $k$ earliest points: \textbf{return} $\{Z_{(i)} \ | \ i \le k\ , Z_{(i)} \in \mathcal{Z} \}$}
\end{algorithmic}
\end{algorithm}

Algorithm~\ref{alg:PPP_order_stats} first draws the number of events in $(a, b]$ from a Poisson distribution. Conditional on the number of events, the event times $Z_i$ are uniformly distributed over $(a, b]$~\cite[par. 4.1]{cox1965theory}. The algorithm returns the order statistics [$Z_{(i)}$], obtained by sorting the event times [$Z_i$] in ascending order. It is necessary to generate all event times to generate the order statistics. Thus, to sample at most $k$ event times we should return the earliest $k$ event times, and line 11 of the Algorithm would be changed to
\begin{center}
\textbf{return} {$\{Z_{(i)} \ | \ i \le k, Z_{(i)} \in \mathcal{Z}\}$}.
\end{center}

The \fct{ppp\_orderstat} function implements constant-rate sampling via the order-statistics algorithm.

\begin{Schunk}
\begin{Sinput}
R> ppp_orderstat(range_t = c(3.14, 6.28), rate = 1/2)
\end{Sinput}
\begin{Soutput}
[1] 3.141663 5.700931
\end{Soutput}
\end{Schunk}

\subsection[Sampling conditional on observing at least m events]{Sampling conditional on observing at least $m$ events}\label{sec:PPP_cond_sampling}

\begin{algorithm}[ht!]
\caption{Sampling with constant intensity $\lambda$ conditional that at least $m$ events occurred in interval $(a, b]$. Relies on generating order statistics analogously to Algorithm~\ref{alg:PPP_order_stats}.}\label{alg:PPP_conditional}
\begin{algorithmic}[1]
\Require $t \in (a, b]$
\State $N \gets N \sim \textrm{TruncatedPoisson}_{N \ge m}\big(\lambda (b-a)\big)$ \Comment{$(m-1)$-truncated Poisson}
\State $t \gets a$
\State $\mathcal{Z} \gets \emptyset$ \Comment{$\mathcal{Z}$ is an ordered set}
\If{N > 0}
    \For{$i \in [N]$}:
        \State $U_i \gets  U_i \sim \textrm{Uniform(0, 1)}$ \Comment{Generate order statistics}
        \State $\mathcal{Z} \gets \mathcal{Z} \cup \{a + (b-a) U_i\} $ 
    \EndFor
    \State $\mathcal{Z} \gets \textrm{sort}(\mathcal{Z})$ 
\EndIf
\State
\Return{$\mathcal{Z}$} \Comment{Up to $k$ earliest points: \textbf{return} $\{Z_{(i)} \ | \ i \le k\ , Z_{(i)} \in \mathcal{Z} \}$}
\end{algorithmic}
\end{algorithm}

Algorithm~\ref{alg:PPP_conditional} is used to generate a point process conditional on observing at least $m$ events. For example, if $\lambda$ is the intensity of tumor generation, it can be used to simulate times of tumor emergence among patients with at least one ($m=1$) tumor. To return the up to $k$ earliest events, we modify line 11 the same way as for Algorithm~\ref{alg:PPP_order_stats}. As an example, in a lifetime simulation we can sample the time of all-cause death by setting in Algorithm~\ref{alg:PPP_conditional} $m=1$, so that at least one event will happen in $(a, b]$, and $k = 1$, to sample only the time of the first event $Z_{(1)}$.

To sample exactly $m$ events, change line 1 of Algorithm~\ref{alg:PPP_conditional} to
\begin{center}
$N \gets m$.
\end{center}

Function \fct{ztppp} simulates times conditional on drawing at least one event - i.e., setting $m=1$ in Algorithm~\ref{alg:PPP_conditional} to sample from a zero truncated Poisson distribution in line 1.
\begin{Schunk}
\begin{Sinput}
R> ztppp(range_t = c(0, 10), rate = 0.001, atmost1 = FALSE)
\end{Sinput}
\begin{Soutput}
[1] 4.411277
\end{Soutput}
\end{Schunk}

Function \fct{ppp\_n} simulates times conditional on drawing exactly $m$ events.
\begin{Schunk}
\begin{Sinput}
R> ppp_n(size = 4, range_t = c(0, 10))
\end{Sinput}
\begin{Soutput}
[1] 1.762014 2.902897 6.751627 9.733794
\end{Soutput}
\end{Schunk}

\section[The general sampling algorithms used in nhppp]{The general sampling algorithms used in \pkg{nhppp}}\label{sec:general-sampling}

The \pkg{nhppp} package uses three well known general sampling algorithms, namely thinning, time transformation or inversion, and order-statistics. These algorithms are efficiently combined to sample from special cases, including cases where the intensity function is a piecewise constant, linear, or log-linear function of time, as described in Section~\ref{sec:sample-nhppp-special}.

The thinning algorithm works with the intensity function $\lambda(t)$, which is commonly available. The inversion and order statistics algorithms have smaller computational cost than the thinning algorithm, but work with the integrated intensity function $\Lambda(t)$ and its inverse $\Lambda^{-1}(z)$, which may not be available. The generic function \fct{draw} is a wrapper function that dispatches to specialized functions depending on the provided arguments. It is useful for general tasks but the specialized functions are probably faster.

\begin{Schunk}
\begin{Sinput}
R> l <- function(t) t
R> L <- function(t) 0.5 * t^2
R> Li <- function(z) sqrt(2 * z)
R> draw(
+    lambda = l, lambda_maj = l(10), range_t = c(5, 10),
+    atmost1 = FALSE, atleast1 = FALSE
+  ) |> head(n = 5)
\end{Sinput}
\begin{Soutput}
[1] 5.179473 5.374814 5.957391 5.992196 6.101935
\end{Soutput}
\begin{Sinput}
R> draw(
+    Lambda = L, Lambda_inv = Li, range_t = c(5, 10),
+    atmost1 = FALSE, atleast1 = FALSE
+  ) |> head(n = 5)
\end{Sinput}
\begin{Soutput}
[1] 5.219264 5.230747 5.369646 5.398531 5.618079
\end{Soutput}
\end{Schunk}

\subsection{The thinning algorithm}\label{sec:thinning}
The thinning algorithm relies on the decomposability of NHPPPs (Section~\ref{sec:properties}) and is described in~\cite{lewis1979thinning}. Let the target NHPPP have intensity function $\lambda(t)$ and $\lambda_*(t) \ge \lambda(t)$ for all $t \in (a, b]$ be a majorizing intensity function. Think of the majorizing function as an easy-to-sample function which is the sum of the intensity of the target point process $\lambda(t)$ and the intensity $\lambda_{reject}(t)$ of its complementary point-process,
$$\lambda_*(t) = \lambda(t) + \lambda_{reject}(t).$$

The acceptance-rejection scheme in Algorithm~\ref{alg:NHPPP_thinning} generates proposal samples with intensity function $\lambda_*(t)$ and stochastically attributes them to the target process (to keep, with probability $\lambda(Z)/\lambda_*(Z)$) or its complement.

\begin{algorithm}[ht!]
\caption{The thinning algorithm for sampling from $\lambda(t)$.}\label{alg:NHPPP_thinning}
\begin{algorithmic}[1]
\Require{
\begin{itemize}
\item[] {$ $}
\item[] { $\lambda_*(t) \ge \lambda(t) \ \forall \ t \in (a, ])$} \Comment{majorizing intensity function}
\item[] {$\mathcal{Z}_* = \{ Z^*_i \ | \ Z^*_i \textrm{ are samples from } \lambda_*(t) \}$} \Comment{$\mathcal{Z}_*$ is an ordered set} 
\end{itemize}}
\State $N \gets |\mathcal{Z}_*|$
\State $\mathcal{Z} \gets \emptyset$ \Comment{$\mathcal{Z}$ is an ordered set}
\If{$N > 0$}
    \For{$i \in [N]$}:
        \State $U_i \gets  U_i \sim \textrm{Uniform(0, 1)}$ 
        \If{$U_i < \lambda(Z^*_{(i)}) / \lambda_*(Z^*_{(i)})$}
            \State $\mathcal{Z} \gets \mathcal{Z} \cup \{ Z^*_{(i)} \} $ 
        \EndIf
    \EndFor
\EndIf
\State
\Return{$\mathcal{Z}$} \Comment{Up to $k$ earliest points: \textbf{return} $\{Z_{(i)} \ | \ i \le k\ , Z_{(i)} \in \mathcal{Z} \}$}
\end{algorithmic}
\end{algorithm}

To sample the earliest $k$ points, one can exit the for loop in lines 4-9 when $k$ events have been sampled in line 7, or, alternatively, return the first up to $k$ points in line 11.

A measure of the efficiency of Algorithm~\ref{alg:NHPPP_thinning} is the proportion of samples that are accepted, which is
\begin{equation}\label{eq:thinning-efficiency}
\frac{\int_a^b{\lambda(t) \textrm{ d}t}}{\int_a^b{\lambda_*(t) \textrm{ d}t}}
\end{equation}
on average. Thus, the closer $\lambda_*(t)$ is to $\lambda(t)$, the more efficient the algorithm.

In practice, $\lambda_*(t)$ can be chosen as one of the special cases in Section~\ref{sec:special_cases}, for which we have fast sampling algorithms. For example, it can be a piecewise constant majorizer. Algorithm~\ref{alg:lambda_majorizer} 
in Appendix~\ref{app:piecewise_majorizer} can automatically generate a piecewise constant majorizer function for intensity functions that are monotonic and possibly non-continuous or Lipschitz continuous and possibly non-monotonic.

The \pkg{nhppp} package has functions that sample from time-varying intensity functions. The first function, \fct{draw\_intensity}, expects a user-provided linear ($\lambda_*(t) = \alpha + \beta t$) or log-linear ($\lambda_*(t) = e^{\alpha + \beta t}$) majorizer function.

\begin{Schunk}
\begin{Sinput}
R> lambda_fun <- function(t) exp(0.02 * t)
R> draw_intensity(
+    lambda = lambda_fun, # linear majorizer
+    lambda_maj = c(intercept = 1.01, slope = 0.03),
+    exp_maj = FALSE, range_t = c(0, 10)
+  ) |> head (n = 5)
\end{Sinput}
\begin{Soutput}
[1] 1.310245 2.094217 2.908682 3.268384 8.007606
\end{Soutput}
\begin{Sinput}
R> draw_intensity(
+    lambda = lambda_fun, # log-linear majorizer
+    lambda_maj = c(intercept = 0.01, slope = 0.03),
+    exp_maj = TRUE, range_t = c(0, 10)
+  ) |> head (n = 5)
\end{Sinput}
\begin{Soutput}
[1] 0.3406743 0.6079479 0.8441584 2.6424551 3.3185387
\end{Soutput}
\end{Schunk}

The second function, \fct{draw\_intensity\_step}, expects a user-provided piecewise linear majorizer
\begin{align*}
    \lambda_*(t) = \begin{cases}
    \lambda_1 &\textrm{ for } t \in [a_1, b_1) = [a, b_1), \\
    \dots &\\
    \lambda_m &\textrm{ for } t \in [a_m, b_m) \textrm{ with } a_{m} = b_{m-1}, \\
    \dots &\\
    \lambda_M &\textrm{ for } t \in [a_M, b_M) = [a_M, b),
    \end{cases}
\end{align*}
which is specified as a vector of length $M+1$ including the points $(a, [b_m]_{m=1}^M)$ and a vector of length $M$ with the values $[\lambda_m]_{m=1}^M$ in each subinterval of $(a, b]$. For example, the following code splits the interval $(0, 10]$ into $M=10$ subintervals of length one. Because \fct{lambda\_fun} is strictly increasing, its value at the upper bound of each subinterval is the supremum of the interval.

\begin{Schunk}
\begin{Sinput}
R> draw_intensity_step(
+    lambda = lambda_fun,
+    lambda_maj_vector = lambda_fun(1:10), # 1:10 (10 intensity values)
+    times_vector = 0:10 # 0:10 (11 interval bounds)
+  ) |> head(n = 5)
\end{Sinput}
\begin{Soutput}
[1] 0.3825378 7.0822941 7.7839779 8.7766992 8.9554954
\end{Soutput}
\end{Schunk}

\subsection{The time transformation or inversion algorithm}\label{sec:inversion}
Algorithm~\ref{alg:NHPPP_inversion} implements the time transformation or inversion algorithm from~\cite{Cinlar1975inversion} and~\citet[par. 4.2]{cox1965theory}. As mentioned in Section~\ref{sec:properties}, strictly monotonic transformations of the carrier space (here, time) of a Poisson point process yield another Poisson Point Process. In equation~\eqref{eq:transform}, choosing the transformation $\tau = u(t) = \Lambda(t)$, so that $\der{u(t)}{t} = \lambda(t)$, results in $\rho(\tau) = 1$.

This means (proof sketched in~\citet[par. 4.2]{cox1965theory}) that we can sample points from a Poisson point process with intensity one over the interval $(\tau_a, \tau_b] = (\Lambda(a), \Lambda(b)]$. Via a similar argument, we transform event times sampled on the transformed scale back to the original scale using $g(t)=\Lambda^{-1}(\tau)$. The transformations $u(\cdot), g(\cdot)$ are not unique -- at least up to the group of affine transformations.

Function \fct{draw\_cumulative\_intensity\_inversion} works with a cumulative intensity function $\Lambda(t)$ and its inverse $\Lambda^{-1}(z)$, if available. If the inverse function is not available (argument \texttt{Lambda\_inv = NULL}), the Brent bisection algorithm is used to invert $\Lambda(t)$ numerically, at a performance cost~\citep{brent-bisection}.

\begin{Schunk}
\begin{Sinput}
R> Lambda_fun <- function(t) 50 * exp(0.02 * t) - 50
R> Lambda_inv_fun <- function(z) 50 * log((z + 50) / 50)
R> draw_cumulative_intensity_inversion(
+    Lambda = Lambda_fun,
+    Lambda_inv = Lambda_inv_fun,
+    range_t = c(5, 10.5),
+    range_L = Lambda_fun(c(5, 10.5))
+  ) |> head(n = 5)
\end{Sinput}
\begin{Soutput}
[1]  6.458937  7.608496  9.060817  9.566278 10.076889
\end{Soutput}
\end{Schunk}

\begin{algorithm}[h!]
\caption{The time transformation or inversion algorithm for sampling given $\Lambda(t), \Lambda^{-1}(z)$~\citep{Cinlar1975inversion, cox1965theory}. The notation $\textrm{PoissonProcess}_1$ indicates sampling event times from a constant rate one Poisson point process.}\label{alg:NHPPP_inversion}
\begin{algorithmic}[1]
\Require{$\Lambda(t), \Lambda^{-1}(z), t \in (a, b]$} \Comment{$\Lambda^{-1}(z)$ possibly numerically}
\State $\tau_a \gets \Lambda(a), \tau_b \gets \Lambda(b)$
\State $\mathcal{C} \gets \mathcal{C} \sim \textrm{PoissonProcess}_{1}(\tau_a, \tau_b)$ \Comment{From Algorithm~\ref{alg:PPP_t} (or~\ref{alg:PPP_conditional} for conditional sampling)} 
\State $\mathcal{Z} \gets \Lambda^{-1}(\mathcal{C})$ \Comment{$\Lambda^{-1}(\cdot)$ as set function, meant elementwise}
\State
\Return{$\mathcal{Z}$} 
\end{algorithmic}
\end{algorithm}

\subsection{The order statistics algorithm}\label{sec:order-stats}
The general order statistics algorithm (Algorithm~\ref{alg:NHPPP_order_stats}) is a direct generalization of Algorithm~\ref{alg:PPP_order_stats}. It first draws the number $N$ of realized events. Conditional on $N$
\begin{equation}\label{eq:nhppp_orderstats1}
\begin{aligned}
U_{(i)} &= \frac{\Lambda(Z_{(i)}) - \Lambda(a)}{\Lambda(b)- \Lambda(a)} \sim \textrm{Uniform}(0,1), \\
Z_{(i)} &= \Lambda^{-1} \Big ( \Lambda(a) + U_{(i)} \big( \Lambda(b)- \Lambda(a) \big) \Big),
\end{aligned}
\end{equation}
as discussed in~\cite{lewis1979thinning}. Algorithm~\ref{alg:NHPPP_order_stats} makes the above explicit.

\begin{algorithm}[h!]
\caption{The order statistics algorithm for sampling from an NHPPP given $\Lambda(t), \Lambda^{-1}(z)$.}\label{alg:NHPPP_order_stats}
\begin{algorithmic}[1]
\Require $\Lambda(t), \Lambda^{-1}(z), t \in (a, b]$ \Comment{$\Lambda^{-1}(z)$ possibly numerically}
\State $N \gets N \sim \textrm{Poisson}\big(\Lambda(b)-\Lambda(a)\big)$
\State $t \gets a$
\State $\mathcal{Z} \gets \emptyset$ \Comment{$\mathcal{Z}$ is an ordered set}
\If{N > 0}
    \For{$i \in [N]$}:
        \State $U_i \gets U_i \sim \textrm{Uniform(0, 1)}$ \Comment{Generate order statistics}
        \State $\mathcal{Z} \gets \mathcal{Z} \cup \{ \Lambda^{-1} \Big( \Lambda(a) + U_i \big( \Lambda(b) - \Lambda(a) \big)\Big) \} $ 
    \EndFor
    \State $\mathcal{Z} \gets \textrm{sort}(\mathcal{Z})$ 
\EndIf
\State
\Return{$\mathcal{Z}$} \Comment{Up to $k$ earliest points: \textbf{return} $\{Z_{(i)} \ | \ i \le k\ , Z_{(i)} \in \mathcal{Z} \}$}
\end{algorithmic}
\end{algorithm}

Sampling up to $k$ earliest points means returning the up to $k$ earliest event times. If $\Lambda(t)$ is a positive linear function of time, $\lambda$ is constant and Algorithm~\ref{alg:NHPPP_order_stats} becomes Algorithm~\ref{alg:PPP_order_stats}.

To sample conditional on observing at least $m$ events in the interval $(a, b]$ see Algorithm~\ref{alg:NHPPP_conditional} in Appendix~\ref{app:conditional_sampling}.
\begin{center}
$N \gets N \sim \mathrm{TruncatedPoisson}_{N \ge m}\big(\Lambda(b) - \Lambda(a)\big)$.
\end{center}

Function \fct{draw\_cumulative\_intensity\_orderstats} works with a cumulative intensity function $\Lambda(t)$ and its inverse $\Lambda^{-1}(z)$, if available.
Function \fct{ztdraw\_cumulative\_intensity} conditions that at least one event is sampled in the interval. As above, if the inverse function is not available (argument \texttt{Lambda\_inv = NULL}), the Brent bisection algorithm is used to invert $\Lambda(t)$ numerically, at a performance cost.

\begin{Schunk}
\begin{Sinput}
R> draw_cumulative_intensity_orderstats(
+    Lambda = Lambda_fun, Lambda_inv = Lambda_inv_fun,
+    range_t = c(4.1, 7.6)
+  )
\end{Sinput}
\begin{Soutput}
[1] 5.091581 5.526070 5.601576 5.762498 6.495684
\end{Soutput}
\end{Schunk}

\begin{Schunk}
\begin{Sinput}
R> ztdraw_cumulative_intensity(
+    Lambda = Lambda_fun, Lambda_inv = Lambda_inv_fun,
+    range_t = c(4.1, 7.6)
+  )
\end{Sinput}
\begin{Soutput}
[1] 5.063676 6.682454 6.749162 6.926164 7.298342
\end{Soutput}
\end{Schunk}

\section{Special cases}\label{sec:special_cases}

The \pkg{nhppp} package implements several special cases where the intensity function $\lambda(\cdot)$, the integrated intensity function $\Lambda(\cdot)$, and its inverse $\Lambda^{-1}(\cdot)$ have straightforward analytical expressions.

\subsection{Sampling a piecewise constant NHPPP}\label{sec:sample-nhppp-pc}

Functions \fct{draw\_sc\_step} and \fct{draw\_sc\_step\_regular} sample piecewise constant intensity functions based on Algorithm~\ref{alg:NHPPP_inversion}. The first can work with unequal-length subintervals $(a_m, b_m]$. The second results in a small computational time improvement when all subintervals are of equal length.
\begin{Schunk}
\begin{Sinput}
R> draw_sc_step(
+    lambda_vector = 1:5, times_vector = c(0.5, 1, 2.4, 3.1, 4.9, 5.9),
+    atmost1 = FALSE, atleast1 = FALSE
+  ) |> head(n = 5)
\end{Sinput}
\begin{Soutput}
[1] 0.8425117 1.3281115 2.3309443 2.6794560 2.7939130
\end{Soutput}
\begin{Sinput}
R> draw_sc_step_regular(
+    lambda_vector = 1:5, range_t = c(0.5, 5.9), atmost1 = FALSE,
+    atleast1 = FALSE
+  ) |> head(n = 5)
\end{Sinput}
\begin{Soutput}
[1] 2.058468 2.100620 2.508954 3.125179 3.604882
\end{Soutput}
\end{Schunk}

Function \fct{vdraw\_sc\_step\_regular} is a vectorized version of \fct{draw\_sc\_step\_regular}. It returns a matrix with one event series per row, and as many columns as the maximum number of events across all draws. 

\begin{Schunk}
\begin{Sinput}
R> vdraw_sc_step_regular(
+    lambda_matrix = matrix(runif(20), ncol = 5), range_t = c(1, 4),
+    atmost1 = FALSE
+  )
\end{Sinput}
\begin{Soutput}
         [,1]     [,2]     [,3]     [,4]
[1,] 2.304123 2.802767       NA       NA
[2,] 2.990953       NA       NA       NA
[3,] 1.840374 2.134357 3.784424 3.816034
[4,] 2.136138 2.703826 3.269631       NA
\end{Soutput}
\end{Schunk}

\noindent \red{The corresponding functions that return at least one event in the interval are \fct{ztdraw\_sc\_step}, \fct{ztdraw\_sc\_step\_regular}, and \fct{vztdraw\_sc\_step\_regular}. }

\subsection{Sampling NHPPPs with linear and log-linear intensities}\label{sec:sample-nhppp-special}

Functions \fct{draw\_sc\_linear} and \fct{ztdraw\_sc\_linear} sample zero or more and at least one event, respectively, from NHPPPs with linear intensity functions. An optional argument (\texttt{atmost1}) returns the first event only.
\begin{align*}
    \lambda(t) =
    \begin{cases}
        \alpha + \beta t &\text{ for } t \in [a, b), t>-\frac{\alpha}{\beta} \\
        0 &\textrm{ otherwise}
    \end{cases}.
\end{align*}

\begin{Schunk}
\begin{Sinput}
R> draw_sc_linear(alpha = 3, beta = -0.5, range_t = c(0, 10)) |> head(n = 5)
\end{Sinput}
\begin{Soutput}
[1] 0.3327657 0.4270154 0.5804320 0.6935027 0.9832093
\end{Soutput}
\begin{Sinput}
R> ztdraw_sc_linear(alpha = 0.5, beta = 0.2, range_t = c(9.999, 10))
\end{Sinput}
\begin{Soutput}
[1] 9.999757
\end{Soutput}
\end{Schunk}

An analogous set of functions (\fct{[nhppp|ztnhppp]\_sc\_loglinear}) samples from log-linear intensity functions

\begin{align*}
    \lambda(t) =
    \begin{cases}
        e^{\alpha + \beta t} &\text{ for } t \in [a, b) \\
        0 &\textrm{ otherwise}
    \end{cases}.
\end{align*}
The sampling algorithm is a variation of Algorithm~\ref{alg:NHPPP_inversion}, as described in~\cite{lewis1976linear}. Example usage follows.
\begin{Schunk}
\begin{Sinput}
R> draw_sc_loglinear(alpha = 1, beta = -0.02, range_t = c(8, 10))
\end{Sinput}
\begin{Soutput}
 [1] 8.028806 8.128887 8.457669 8.483558 8.498647 8.503109 8.522725
 [8] 8.665979 8.671737 8.978065 8.981105 9.493691 9.815000 9.909167
\end{Soutput}
\end{Schunk}

\begin{Schunk}
\begin{Sinput}
R> ztdraw_sc_loglinear(alpha = 1, beta = -0.02, range_t = c(9, 10))
\end{Sinput}
\begin{Soutput}
[1] 9.038160 9.075722 9.238302
\end{Soutput}
\end{Schunk}

\section[Comparisons with other R packages]{Comparisons with other \proglang{R} packages}\label{sec:other-R-packages}

Table~\ref{tab:R-packages} lists five \proglang{R} packages that simulate from NHPPPs, including \pkg{nhppp}. \red{We did not consider research code that is not an \proglang{R} package in the Comprehensive R Archive Network or is developed in other languages. For example, we do not run comparisons with the \proglang{R} and \proglang{Python} code for sampling from piecewise constant NHPPPs with regular time intervals in~\cite{garibay2024nps}. (Their code corresponds to the \fct{vdraw\_sc\_step\_regular} function in \pkg{nhppp}.)}

Package \pkg{reda}~\citep{reda-package} focuses on recurrent event data analysis and can simulate NHPPPs with the inversion and thinning algorithms using the \fct{simEvent} function. It can take function object arguments for $\lambda(t)$. When using the thinning algorithm, it takes a constant majorizer. For the inversion algorithm, it approximates $\Lambda(t)$ and its inverse numerically, at a computational cost.

Package \pkg{simEd}~\citep{simEd-package} includes various functions for simulation education. Function \fct{thinning} implements the homonymous algorithm for drawing points from an NHPPP. Users can specify the intensity function and a piecewise constant or linear majorizer function.

Package \pkg{IndTestPP}~\citep{IndTestPP-package} provides a framework for exploring the dependence between two or more realizations of point processes. It includes the ancillary function~\fct{simNHPc} for simulating NHPPPs with the inversion or thinning algorithms. The function's argument is a piecewise constant approximation of the intensity function via a vector of evaluations, each corresponding to unit length subintervals. This resolution may not be adequate to simulate processes that change fast over a unit time interval.

Package \pkg{NHPoisson}~\citep{NHPoisson-jss, NHPoisson-package} fits NHPPP models to data and is not really geared towards mathematical simulation. Its \fct{simNHP.fun} function provides the ability for simulation-based inference via an implementation of the inversion algorithm. This function is designed to work with the package's inference machinery and is not practical to use for simulation, because the user has no direct control over the function's rescaling of the time axis.

The claimed advantage of \pkg{nhppp} over the existing packages is that
\begin{itemize}
\item it samples from the target NHPPP and not from a numerical approximation thereof, e.g., as \pkg{IndTestPP} does.
\item It can  sample conditional on observing at least one event in the interval, which no other package implement.
\item It accepts user-provided random number stream objects, which is useful for implementing simulation variance reduction techniques such as common random numbers~\citep{wright1979crn} and antithetic variates~\citep{hammersley1956av}.
\item It is fast and memory efficient, \red{both for the non-vectorized functions that are implemented in native \proglang{R} and for the vectorized functions that use \proglang{C++} plugins via the \pkg{Rcpp} package~\citep{rcpp-package}.} \pkg{nhppp} has specialized functions to leverage additional information about the point process, such as $\Lambda(t), \Lambda^{-1}(z)$, when available, which can result in faster simulation  use the cumulative intensity function and its inverse, often at a computational speed advantage. 
\end{itemize}

\begin{sidewaystable}
\caption{\textbf{NHPPP generation in \proglang{R} packages}} \label{tab:R-packages}
\vspace{5pt}
\footnotesize
\centering
\begin{tabular}{lllllllll}
\toprule
\textbf{\proglang{R} package} &
\textbf{Function} &
\multicolumn{3}{c}{\textbf{Algorithms (inputs)}} & 
\begin{tabular}[c]{@{}l@{}}\textbf{Sample only} \\\textbf{earliest event}\end{tabular} & 
\begin{tabular}[c]{@{}l@{}}\textbf{Custom} \\\textbf{RNG}\end{tabular}& 
\begin{tabular}[c]{@{}l@{}}\textbf{Simulate} \\\textbf{given} $N>0$\end{tabular}&
\begin{tabular}[c]{@{}l@{}}\textbf{Vectorized} \\\textbf{functions}\end{tabular}\\
\cmidrule(lr){3-5} 
  &  & \textbf{Thinning} & \textbf{Inversion} & \begin{tabular}[c]{@{}l@{}}\textbf{Order} \\\textbf{statistics} \end{tabular}& & &\\
\midrule
\pkg{nhppp} &
[see text] &
$\lambda(t)$, $\lambda_*(t)$ &
$\Lambda(t), \Lambda^{-1}(z)$ &
$\Lambda(t), \Lambda^{-1}(z)$ &
Yes &  
\begin{tabular}[c]{@{}l@{}}\pkg{rstream} \\objects\end{tabular} & 
Yes &
\begin{tabular}[c]{@{}l@{}}For piecewise \\constant\\ intensity\end{tabular}\\
\addlinespace[1em]
\pkg{reda} & 
\fct{simEvent}   &
$\lambda(t)$, $\lambda_{*}$ constant   &
\begin{tabular}[c]{@{}l@{}}$\lambda(t)$ \\(no $\Lambda(t), \Lambda^{-1}(z)$)\end{tabular} &
No &
Yes &
No &
No &
No \\
\addlinespace[1em]
\pkg{simEd} & 
\fct{thinning}  &
$\lambda(t)$, $[\lambda_{*m}]_{m=1}^M$ &
No & 
No &
No &
No &
No &
No \\
\addlinespace[1em]
\pkg{IndTestPP} &
\fct{simNHPc} &
$[\lambda_m]_{m=1}^M$, $\lambda_{*}$ constant  & 
\begin{tabular}[c]{@{}l@{}}$[\lambda_m]_{m=1}^M$ \\(no $\Lambda(t), \Lambda^{-1}(z)$) \end{tabular}&
No &
No & 
No & 
No &
No \\
\addlinespace[1em]
\pkg{NHPoisson} & 
\fct{simNHP.fun} &
No &
\begin{tabular}[c]{@{}l@{}}$\lambda(t)$, \\(no $\Lambda(t), \Lambda^{-1}(z)$)\end{tabular} &
No &
No &
No &
No &
No \\
\bottomrule
\end{tabular}
\begin{flushleft}
RNG: random number generator object.
\end{flushleft}
\end{sidewaystable}

\newpage
\section{Illustrations} \label{sec:illustrations}

Depending on the application, we may have access to the intensity function
or the integrated intensity function.
We compared the \proglang{R} packages in Table~\ref{tab:R-packages} for sampling from a non-monotonic and highly non-linear intensity function for which the integrated intensity function can be derived analytically.

\subsection{The target NHPPP to be simulated}\label{sec:illustration-target}

Consider the example
\begin{equation}\label{eq:illustration}
\begin{aligned}
\lambda(t) &= e^{rt}(1+\sin wt), \\
\Lambda(t) &= \dfrac{e^{rt}(r\sin wt - w\cos wt) + w}{r^2+w^2}+\dfrac{e^{rt}-1}{r} %- \dfrac{r^2-r+1}{r^3+r}
\end{aligned}
\end{equation}
of a sinusoidal intensity function $\lambda(t)$ scaled to have an exponential amplitude and one of its antiderivatives $\Lambda(t)$, with such a constant term that $\Lambda(0)=0$.  For the numerical study we set $r=0.2$, $w=1$, and $t \in (0, 6\pi]$. There is no  analytic inverse function for this example.
However, we can precompute \fct{Li}, a good numerical approximation to $\Lambda^{-1}(z)$. We will use it in Section~\ref{sec:example-time-performance} to compare the time performance of functions that use the inversion and order statistics algorithms when $\Lambda^{-1}$ is available versus not.
\begin{Schunk}
\begin{Sinput}
R> l <- function(t) (1 + sin(t)) * exp(0.2 * t)
R> L <- function(t) {
+    exp(0.2 * t) * (0.2 * sin(t) - cos(t)) / 1.04 +
+      exp(0.2 * t) / 0.2 - 4.038462
+  }
R> Li <- approxfun(
+    x = L(seq(0, 6 * pi, 10^-3)),
+    y = seq(0, 6 * pi, 10^-3), rule = 2
+  )
\end{Sinput}
\end{Schunk}

\begin{Schunk}
\begin{figure}

\caption[The $\lambda(t)$ (left) and $\Lambda(t)$ used in the illustration]{{\bf The $\lambda(t)$ (left) and $\Lambda(t)$ used in the illustration.}}\label{fig:example-function-plot}
\includegraphics[width=\maxwidth]{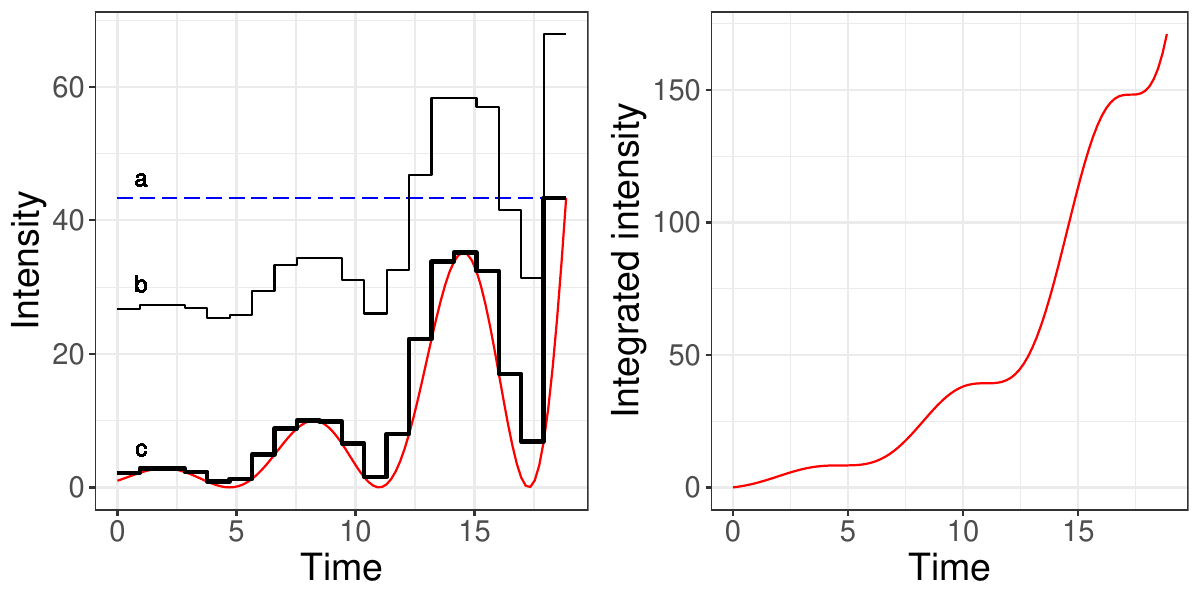} 
\begin{flushleft}
Also shown three majorizing functions (left panel, marked a, b, c) that are used with the thinning algorithm in the analyses.
\end{flushleft}
\end{figure}
\end{Schunk}

Fig~\ref{fig:example-function-plot} graphs the intensity function and three majorizing functions over the interval of interest, which will be needed for the thinning algorithm.

The first, $\lambda_{*a}(t) = 43.38$, shown as a dashed blue line, is is a constant majorizer equal to the maximum of the intensity function. A constant majorizer may be a practical choice when only an upper bound is known for $\lambda(t)$.
From~\eqref{eq:thinning-efficiency}, the \red{analytic} efficiency of the thinning algorithm using this majorizer is
$0.209$.

The second, $\lambda_{*b}(t)$, shown as a thin black line, is a piecewise constant envelope generated automatically from Algorithm~\ref{alg:lambda_majorizer} 
(in Appendix~\ref{app:piecewise_majorizer}) with $20$ equal-length subintervals and Lipschitz cone coefficient ${K = 52.05}$. We set $K$ equal to the maximum value of $|\der{\lambda(t)}{t}|$ in the interval, attained at $6\pi$.
The \red{analytic} efficiency of the thinning algorithm using this majorizer is
$0.245$.

The third, $\lambda_{*c}(t)$, shown as a thicker black line, is a tighter piecewise constant majorizer with the same $20$ equal-length subintervals that is constructed by finding a least upper bound in each subinterval. The \red{analytic} efficiency of the thinning algorithm with the third
majorizer is
$0.718$.

\subsection{Simulation functions and algorithms}\label{sec:methods-sim}

We sampled series of events from the target NHPPP using the packages and functions listed in Table~\ref{tab:R-packages}. We repeated the sampling $\ensuremath{10^{4}}$ times, recording all simulated points (event times). We also recorded the median computation time for drawing one series of events with single-threaded computation on modern hardware.

From the \pkg{nhppp} package we use
\begin{enumerate}
\item  two functions that take as argument the intensity function and are based on Algorithm~\ref{alg:NHPPP_thinning} (thinning): \fct{draw\_intensity}, which uses linear majorizers such as $\lambda_{*a}$, and \fct{draw\_intensity\_step}, which uses piecewise constant majorizers such as $\lambda_{*b}$ and $\lambda_{*c}$ in the example.
\item  Function \fct{draw\_cumulative\_intensity\_inversion}, which takes as argument the cumulative intensity function $\Lambda(t)$ and is based on Algorithm~\ref{alg:NHPPP_inversion} (time transformation/inversion), and
\item  function \fct{draw\_cumulative\_intensity\_orderstats}, which also uses $\Lambda(t)$ and is based on Algorithm~\ref{alg:NHPPP_order_stats} (order statistics).
\end{enumerate}

Regarding the other \proglang{R} packages in Table~\ref{tab:R-packages}, we used all except for \pkg{NHPoisson}, whose simulation function is tailored to supporting simulation based inference for data analysis and is not practical to use as a standalone function.
(Its implementation does not allow the user to control the scaling of the time axis in a practical way.)
However, its source code/algorithm is very similar to that of the \pkg{IndTestPP} simulation function, which is developed by the same authors.

We used the metrics in Table~\ref{tab:sim_metrics} to assess simulation performance with each function. We compared the empirical versus the simulated distributions of number of events and event times over $J = 100$ simulation runs.

\subsection{Simulation performance with respect to number of events}\label{sec:sim-counts}

We calculated the absolute and relative bias in the first two moments of the empirical distribution in the counts of events, the bounds of equal-tailed confidence intervals at the 95, 90, 75, and 50 percent levels, a $\chi^2$-distributed goodness of fit statistic and its $p$-value, and the Wasserstein-1 distance $W_1$ between the empirical and the theoretical count distributions and the asymptotic one sided $p$~value to reject whether $W_1 = 0$ according to~\cite{sommerfeld2018inference}. $W_1$ is the smallest mass that has to be redistributed so that one distribution matches the other. $W_1$ is equal to the unsigned area between the cumulative distribution functions of the compared distributions. For example, $W_1 = 5.25$ means that the mass that must be moved to transform one density to the other is no less than $5.25$ counts and a $W_1 = 0$ implies perfect fit.

\begin{table}
\caption{\textbf{Simulation metrics for the number of counts}}
\label{tab:sim_metrics}
\vspace{5pt}
\small
\centering
\begin{tabular}{lll}
\toprule
\textbf{Metric} &
\textbf{Definition} &
\textbf{Description} \\
\midrule
Bias in mean & 
$B_\mu = \frac{1}{J}\sum_j{n_j} - N$ & 
\begin{tabular}[c]{@{}l@{}}
Mean difference from target in\\
the number of counts. 
\end{tabular} \\
\addlinespace
Relative bias in mean & 
$B_{\mu,rel} = \frac{B_\mu}{N}$ & 
\begin{tabular}[c]{@{}l@{}}
Mean proportional difference\\
from target in the number of\\
counts.
\end{tabular} \\
\addlinespace
Bias in variance & 
$B_V = \frac{1}{J}\sum_j{ \big( n_j - \frac{1}{J} \sum_j {n_j} \big)^2} - V$ & 
\begin{tabular}[c]{@{}l@{}}
Mean difference from target\\ 
in variance of counts. 
\end{tabular} \\
\addlinespace
Relative bias in variance & 
$B_{V, rel} = \frac{B_V}{V}$ & 
\begin{tabular}[c]{@{}l@{}}
Mean proportional difference\\ 
from target in variance of\\ 
counts. 
\end{tabular} \\
\addlinespace
\begin{tabular}[c]{@{}l@{}} Equal-tailed $p$\% \\confidence interval bounds\end{tabular} & 
$n_{[p/2]}, n_{[1-p/2]}$ & 
\begin{tabular}[c]{@{}l@{}}
Quantiles of the empirical\\ 
distribution of counts.
\end{tabular} \\
\addlinespace
Goodness of fit $p$~value & 
Statistic $\sum_x \frac{(O_x - E_x)^2}{E_x} \sim \chi^2_{U-L+1}  $ & 
\begin{tabular}[c]{@{}l@{}} 
Left-tail $p$~value.\\
$p$~values near 1 imply good fit.
\end{tabular} \\
\addlinespace
Wasserstein-1 distance& 
\begin{tabular}[c]{@{}l@{}}
$W_1$, the smallest rearrangement \\
of probability mass so that one\\
distribution matches the other.   
\end{tabular} & 
$W_1 = 0$ implies good fit \\
\addlinespace
$p$~value for $W_1\ne0$ &  
Asymptotic theory $p$~value & 
\begin{tabular}[c]{@{}l@{}} 
Two-sided $p$~value.\\
$p$~values near 1 imply good fit.
\end{tabular} \\
\bottomrule
\end{tabular}
\begin{flushleft}
In the Table, $j \in [J]$ indexes simulations, $n_j$ is the number of counts in simulation $j$, $N = \Lambda(6\pi)-\Lambda(0)$ is the theoretical mean number of counts, and $V = \Lambda(6\pi)-\Lambda(0)=N$ the theoretical variance. The lower and upper bounds of an equal-tailed $p$\% confidence interval, $p \in \{95, 90, 75, 50\}$, are denoted with $n_{[p/2]}, n_{[1-p/2]}$, respectively. 
For the goodness of fit, we created bins $[0, L), [L, L+1), \dots, [U, \infty)$, where $L, U$ are the $0.001$ and $0.999$ percentiles of the Poisson distribution with parameter $\Lambda(6\pi) - \Lambda(0)$. 
We indexed bins with $x \in \{ 1, \dots, U-L+2\}$. 
%In the example, $L = \Sexpr{stats::qpois(0.001, L(6*pi)-L(0))}$ and $U = \Sexpr{stats::qpois(0.999, L(6*pi)-L(0))}$. 
The goodness of fit statistic contrasts the observed ($O_x$) versus expected ($E_x$) numbers of events over the bins and it is compared with a $\chi^2_{U-L+1}$ distribution to obtain a $p$~value.
\end{flushleft}
\end{table}

The results for the~\pkg{nhppp} functions in Fig~\ref{fig:ecdf-nhppp-pkg-counts} and Table~\ref{tab:nhppp-results-counts} suggest excellent simulation performance.

\begin{Schunk}
\begin{figure}[ht!]
\caption{{\bf Theoretical (red) and empirical (black) cumulative distribution functions for event counts in the illustration example with \pkg{nhppp} functions. }}\label{fig:ecdf-nhppp-pkg-counts}
\includegraphics[width=\maxwidth]{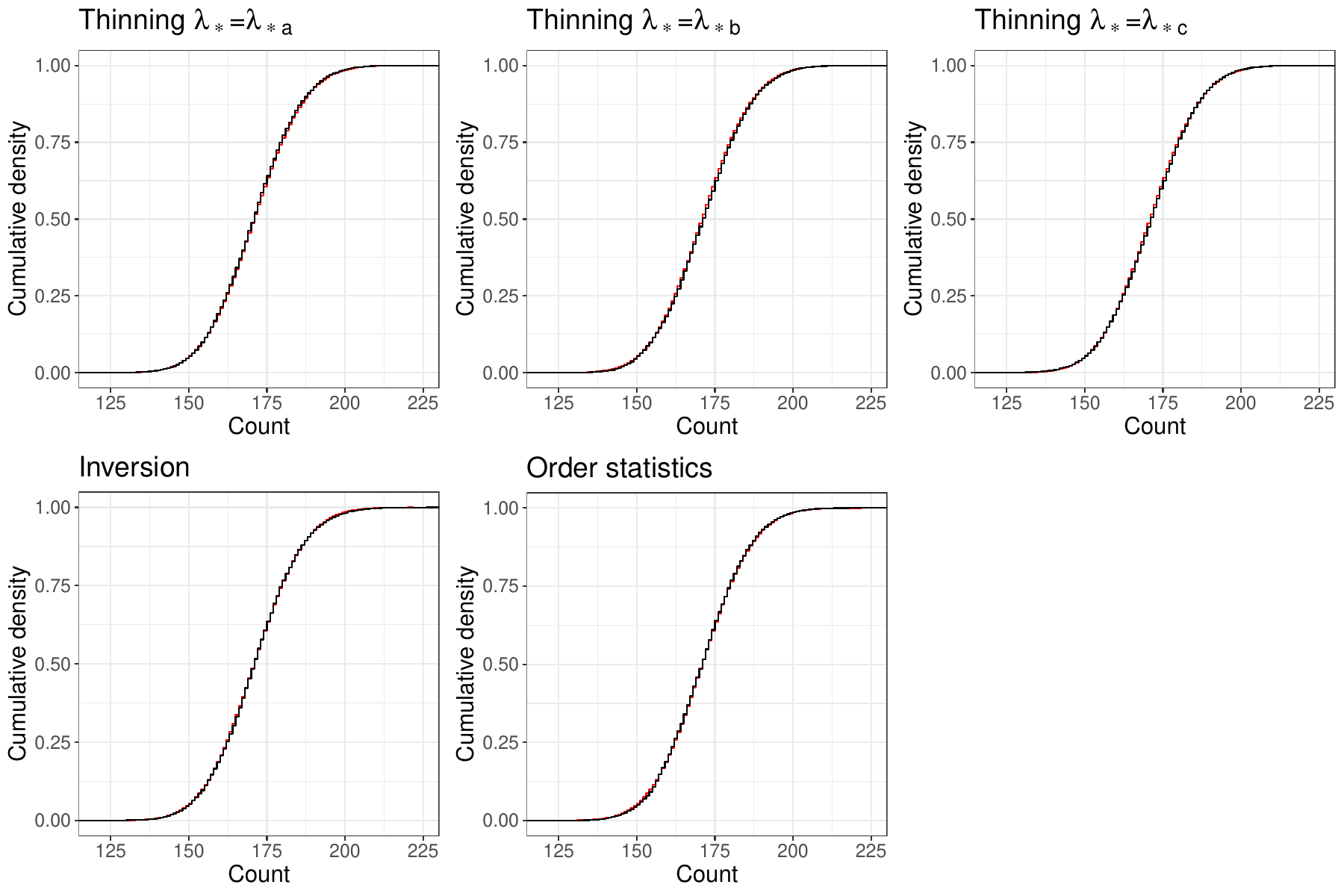} 
\begin{flushleft}
The unsigned area between the theoretical and empirical curves equals the Wasserstein-1 distance in Table~\ref{tab:nhppp-results-counts}.
\end{flushleft}
\end{figure}
\end{Schunk}

% latex table generated in R 4.3.1 by xtable 1.8-4 package
% Fri Feb  2 21:22:31 2024
\begin{table}[ht!]
\caption{\textbf{Simulated total number of events with \pkg{nhppp} functions for the illustration example.}}\label{tab:nhppp-results-counts}
\vspace{5pt}
\centering
\scalebox{0.75}{
\begin{tabular}{rlllll}
  \hline
 & Thinning $\lambda_*$=$\lambda_{*a}$ & Thinning $\lambda_*$=$\lambda_{*b}$ & Thinning $\lambda_*$=$\lambda_{*c}$ & Inversion & Order statistics \\ 
  \hline
Sample mean & 171.057 & 171.257 & 171.322 & 171.193 & 171.131 \\ 
  $B_\mu$ & -0.078 & 0.122 & 0.187 & 0.058 & -0.004 \\ 
  $B_{\mu, rel}$ & -0.045 & 0.071 & 0.109 & 0.034 & -0.002 \\ 
  Sample variance & 175.015 & 168.218 & 173.918 & 166.950 & 166.933 \\ 
  $B_V$ & 3.880 & -2.917 & 2.783 & -4.185 & -4.201 \\ 
  $B_{V, rel}$ & 2.267 & -1.704 & 1.626 & -2.445 & -2.455 \\ 
  Goodness of fit, $\chi^2$ [$p$~value] & 0.145 [1.000] & 0.160 [1.000] & 0.117 [1.000] & 0.384 [1.000] & 0.229 [1.000] \\ 
  $W_1$ [$p$~value] & 0.194 [1.000] & 0.189 [1.000] & 0.231 [0.997] & 0.195 [1.000] & 0.187 [1.000] \\ 
  Equal tail 95\% CI = [146, 197] & [146, 197] & [146, 197] & [146, 197] & [146, 197] & [146, 197] \\ 
  Equal tail 90\% CI = [150, 193] & [150, 193] & [150, 193] & [150, 193] & [150, 193] & [150, 193] \\ 
  Equal tail 75\% CI = [156, 186] & [156, 186] & [156, 186] & [156, 187] & [156, 186] & [156, 186] \\ 
  Equal tail 50\% CI = [162, 180] & [162, 180] & [162, 180] & [162, 180] & [162, 180] & [162, 180] \\ 
   \hline
\end{tabular}
} %scalebox
\begin{flushleft}
Equal tail $p$\% CI: a confidence interval whose bounds are the $p/2$ and $(1-p/2)$ count percentiles of the respective cumulative distribution function.
\end{flushleft}
\end{table}

The respective results for the \proglang{R} packages are in Fig~\ref{fig:r-pkgs-pkg-counts} and Table~\ref{tab:r-packages-results-counts1}. The simulation performance with the \pkg{reda} functions is excellent. Performance with \pkg{simEd} and \pkg{IndTestPP} functions depends on the adequacy with which they approximate the target density. In this example, the approximation accuracy is not ideal for either package, but is somewhat worse for \pkg{IndTestPP}.

\begin{Schunk}
\begin{figure}[ht!]
\caption{{\bf Theoretical (red) and empirical (black) cumulative distribution functions for event counts in the illustration example with the \proglang{R} packages in Table~\ref{tab:R-packages}.}}\label{fig:r-pkgs-pkg-counts}
\includegraphics[width=\maxwidth]{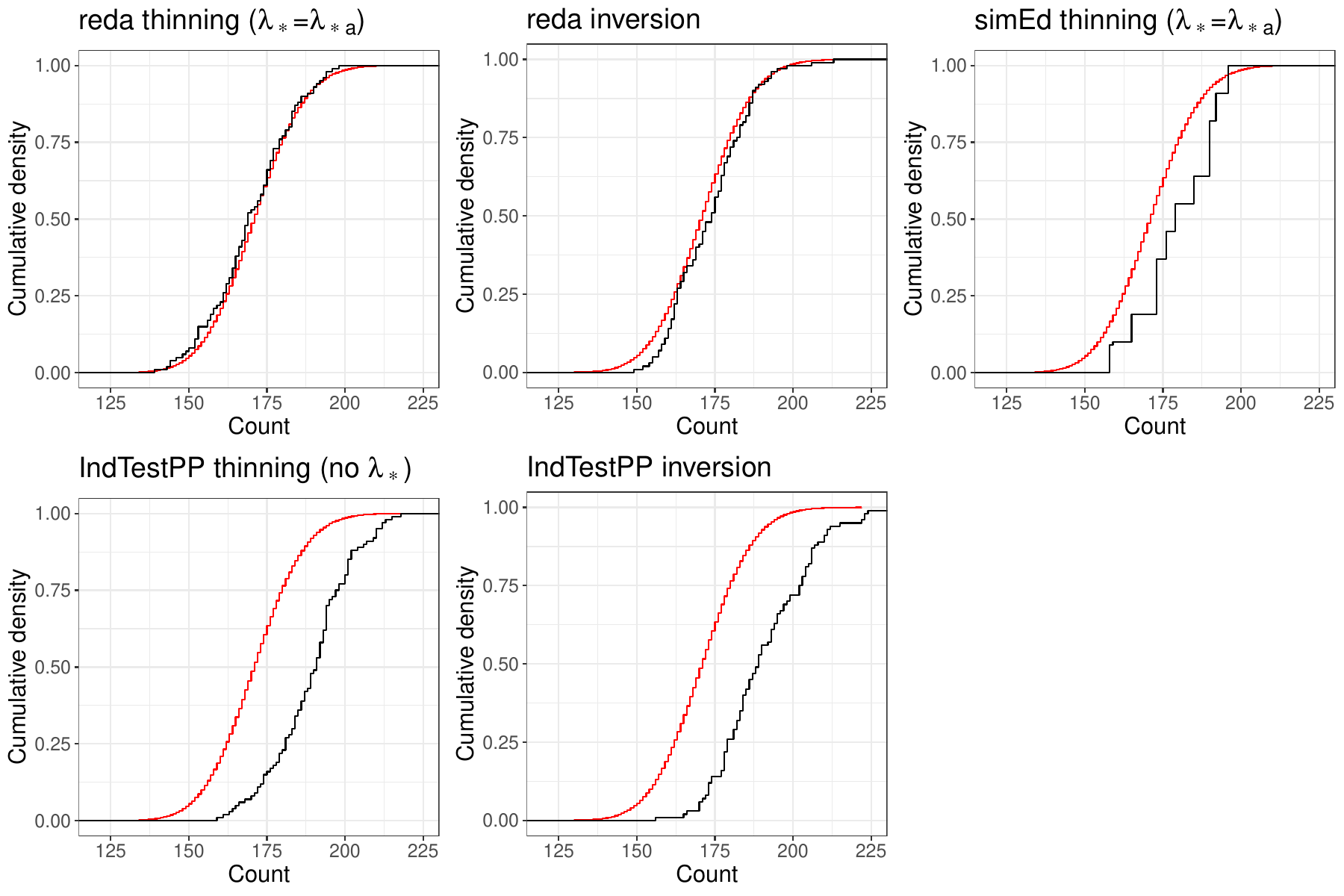} 
\begin{flushleft}
The unsigned area between the theoretical and empirical curves equals the Wasserstein-1 distance in Table~\ref{tab:nhppp-results-counts}.
\end{flushleft}
\end{figure}
\end{Schunk}

% latex table generated in R 4.3.1 by xtable 1.8-4 package
% Fri Feb  2 21:22:32 2024
\begin{table}[H]
\caption{\textbf{Simulated total number of events with the \proglang{R} packages of Table~\ref{tab:R-packages} for the illustration example.}}\label{tab:r-packages-results-counts1} 
\vspace{5pt}
\centering
\scalebox{0.75}{
\begin{tabular}{rp{1in}p{1in}p{1in}p{1in}p{1.1in}}
  \hline
 & \pkg{reda} thinning, $\lambda_*$=$\lambda_{*a}$ & \pkg{reda} inversion & \pkg{simEd} thinning, $\lambda_*$=$\lambda_{*a}$ & \pkg{IndTestPP} thinning, no $\lambda_*$ & \pkg{IndTestPP} inversion \\ 
  \hline
Sample mean & 172.430 & 174.170 & 179.910 & 190.490 & 191.030 \\ 
  $B_\mu$ & 1.295 & 3.035 & 8.775 & 19.355 & 19.895 \\ 
  $B_{\mu, rel}$ & 0.757 & 1.774 & 5.128 & 11.310 & 11.626 \\ 
  Sample variance & 168.429 & 145.193 & 155.355 & 194.838 & 191.484 \\ 
  $B_V$ & -2.705 & -25.942 & -15.779 & 23.704 & 20.349 \\ 
  $B_{V, rel}$ & -1.581 & -15.159 & -9.220 & 13.851 & 11.891 \\ 
  Goodness of fit, $\chi^2$ [$p$~value] & 6.830 [1.000] & 10.720 [1.000] & 67.482 [0.994] & 226.107 [<0.001] & 237.199 [<0.001] \\ 
  $W_1$ [$p$~value] & 1.453 [0.256] & 3.083 [0.112] & 8.856 [<0.001] & 19.356 [0.086] & 19.896 [0.170] \\ 
  Equal tail 95\% CI = [146, 197] & [152, 199] & [152, 196] & [161, 203] & [163, 214] & [168, 217] \\ 
  Equal tail 90\% CI = [150, 193] & [154, 195] & [154, 192] & [161, 203] & [167, 214] & [170, 215] \\ 
  Equal tail 75\% CI = [156, 186] & [158, 189] & [161, 187] & [169, 202] & [174, 205] & [176, 207] \\ 
  Equal tail 50\% CI = [162, 180] & [162, 181] & [165, 180] & [170, 183] & [178, 201] & [181, 200] \\ 
   \hline
\end{tabular}
}
\begin{flushleft}
Equal tail $p$\% CI: a confidence interval whose bounds are the $p/2$ and $(1-p/2)$ count percentiles of the respective cumulative distribution function.    
\end{flushleft}
\end{table}

\subsection{Event times}\label{sec:example-event-times}

We compared the theoretical and empirical distribution of event times for all $J=\ensuremath{10^{4}}$ event time draws. We calculated a goodness of fit statistic by binning realized times in $70$ bins and its $p$~value, by comparing the statistic against the $\chi^2_{69}$ distribution. We also calculated the $W_1$ distance between these distributions and its associated $p$~value.

\begin{figure}[H]
\caption{{\bf Simulated event times with \pkg{nhppp}.}}\label{fig:epdf-nhppp-pkg-times}
{\centering 
\includegraphics[width=\maxwidth,height=\textheight,keepaspectratio=true]{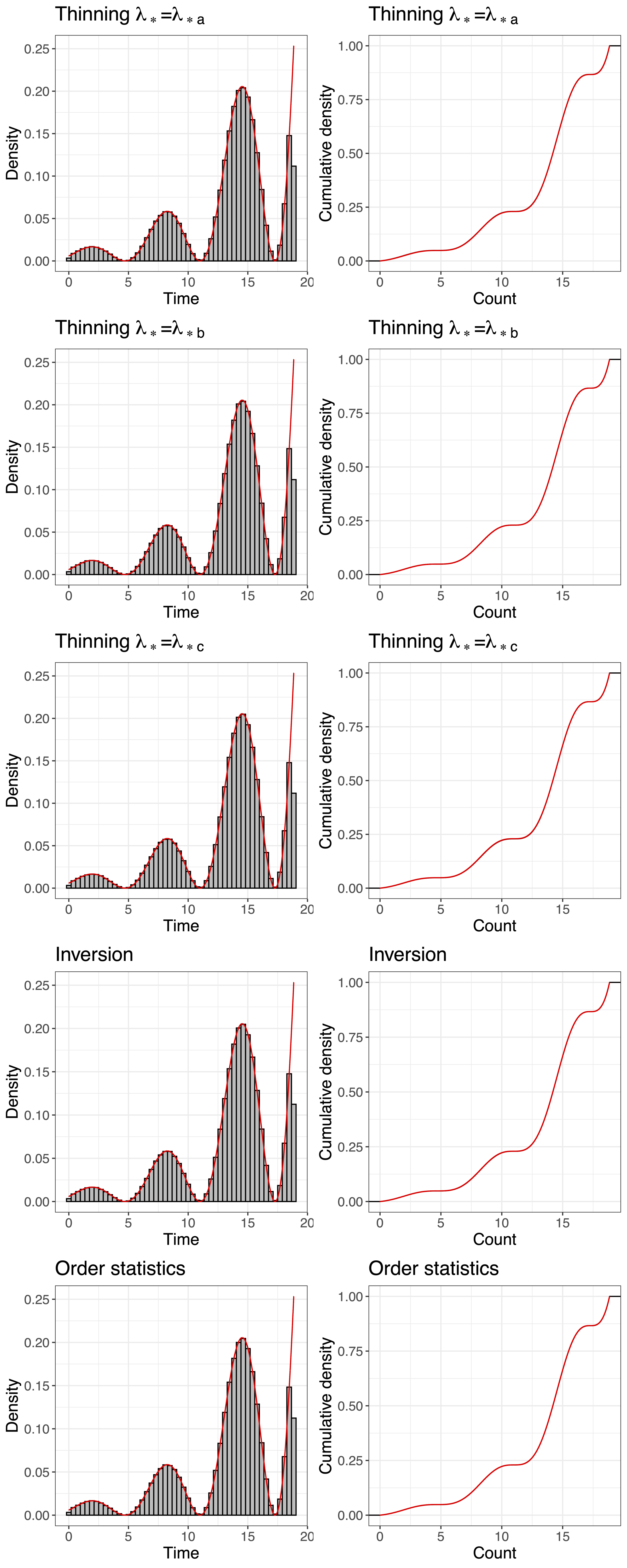} 
}
\begin{flushleft}
Left column: histogram (gray) and theoretical distribution (red) of event times; right column: empirical (black) and theoretical (red) cumulative distribution function. The unsigned area between the empirical and cumulative distribution functions is the $W_1$ distance in Table~\ref{tab:nhppp-gof-times}.
\end{flushleft}
\end{figure}

Fig~\ref{fig:epdf-nhppp-pkg-times} and Table~\ref{tab:nhppp-gof-times} indicate excellent simulation performance with the \pkg{nhppp} functions.

% latex table generated in R 4.3.1 by xtable 1.8-4 package
% Fri Feb  2 21:32:56 2024
\begin{table}[ht!]
\centering
\caption{\textbf{Goodness of fit of simulated event times with \pkg{nhppp} functions for the example.}} 
\vspace{5pt}
\label{tab:nhppp-gof-times}
\scalebox{1.0}{
\begin{tabular}{rll}
  \hline
 & Goodness of fit, $\chi^2$ [$p$~value] & $W_1$ [$p$~value] \\ 
  \hline
Thinning $\lambda_*$=$\lambda_{*a}$ & 0.004 [1.000] & 0.396 [1.000] \\ 
  Thinning $\lambda_*$=$\lambda_{*b}$ & 0.004 [1.000] & 0.361 [1.000] \\ 
  Thinning $\lambda_*$=$\lambda_{*c}$ & 0.004 [1.000] & 0.338 [1.000] \\ 
  Inversion & 0.004 [1.000] & 0.347 [1.000] \\ 
  Order statistics & 0.004 [1.000] & 0.350 [1.000] \\ 
   \hline
\end{tabular}
}
\end{table}

\newpage
%\begin{Schunk}
%\end{Schunk}

%\begin{Schunk}
\begin{figure}[H]
\caption{{\bf Simulated event times with the \proglang{R} packages in Table~\ref{tab:R-packages}.}}\label{fig:epdf-r-pkgs-times}
{\centering \includegraphics[width=\maxwidth,height=\textheight,keepaspectratio=true]{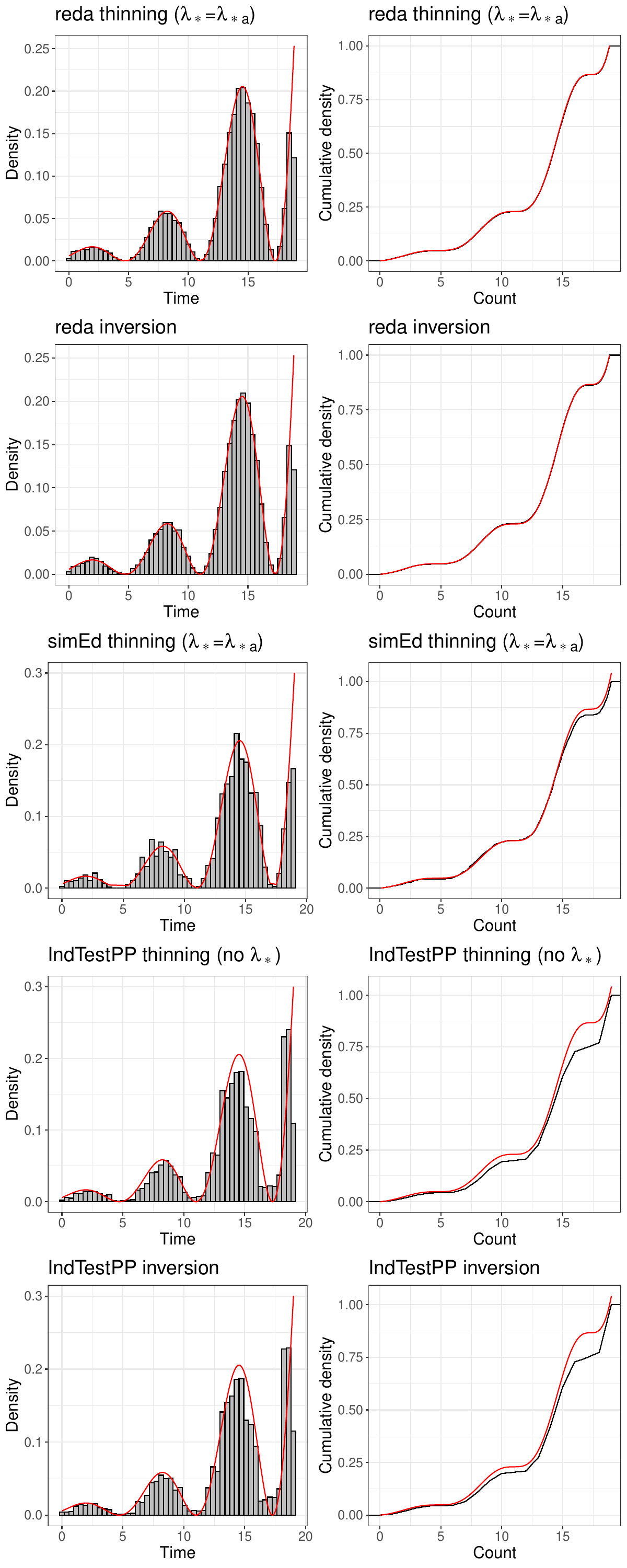} 
}
\begin{flushleft}
Left column: histogram (gray) and theoretical distribution (red) of event times; right column: empirical (black) and theoretical (red) cumulative distribution function. The unsigned area between the empirical and cumulative distribution functions is the $W_1$ distance in Table~\ref{tab:r-pkgs-gof-times}.
\end{flushleft}
\end{figure}
%\end{Schunk}

Fig~\ref{fig:epdf-r-pkgs-times} and Table~\ref{tab:r-pkgs-gof-times} indicate excellent simulation performance with the \pkg{reda} functions. The simulation performance with the \pkg{simEd} and \pkg{IndTestPP} functions, which rely on approximations, is not as good.

% latex table generated in R 4.3.1 by xtable 1.8-4 package
% Fri Feb  2 21:32:56 2024
\begin{table}[ht!]
\caption{\textbf{Goodness of fit of simulated event times with \proglang{R} functions in Table~\ref{tab:R-packages}.}}
\vspace{5pt}
\label{tab:r-pkgs-gof-times}
\centering
\scalebox{1.0}{
\begin{tabular}{rll}
  \hline
 & Goodness of fit, $\chi^2$ [$p$~value] & $W_1$ [$p$~value] \\ 
  \hline
\pkg{reda} thinning ($\lambda_*$=$\lambda_{*a}$) & 0.012 [1.000] & 0.356 [1.000] \\ 
  \pkg{reda} inversion & 0.010 [1.000] & 0.354 [1.000] \\ 
  \pkg{simEd} thinning ($\lambda_*$=$\lambda_{*a}$) & 0.028 [1.000] & 0.338 [0.990] \\ 
  \pkg{IndTestPP} thinning (no $\lambda_*$) & 0.460 [1.000] & 2.152 [0.930] \\ 
  \pkg{IndTestPP} inversion & 0.490 [1.000] & 2.372 [0.927] \\ 
   \hline
\end{tabular}
}
\end{table}

\subsection{Time performance}\label{sec:example-time-performance}

\subsubsection{Time performance of non-vectorized functions}\label{sec:nonvectorized-functions}

To indicate time performance, we benchmarked functions by recording execution times when drawing a series of points (Fig~\ref{fig:comptimes_all_samples}). We also benchmarked functions for drawing the first-occurring event, because \pkg{nhppp} functions can sample the first time more efficiently when the inversion algorithm is used (Fig~\ref{fig:comptimes_one_sample}).

\begin{Schunk}
\begin{figure}[ht!]
\caption[Computation times when drawing all events in interval]{\textbf{Computation times when drawing all events in interval.}}\label{fig:comptimes_all_samples}
\includegraphics[width=\maxwidth]{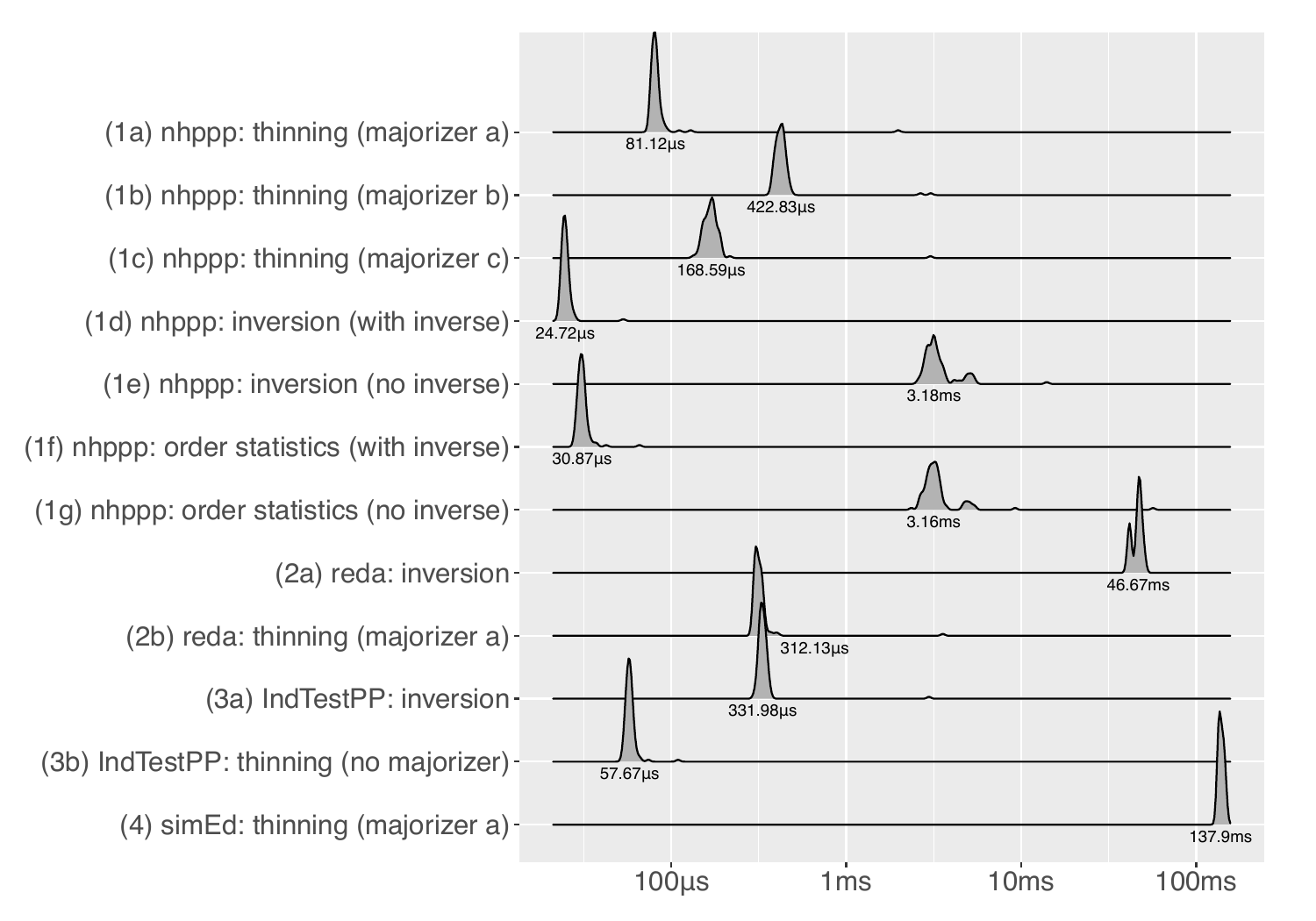} 
\end{figure}
\end{Schunk}

\begin{Schunk}
\begin{figure}[ht!]
\caption[Computation times when drawing the first event in interval]{{\bf Computation times when drawing the first event in interval.}}\label{fig:comptimes_one_sample}
\includegraphics[width=\maxwidth]{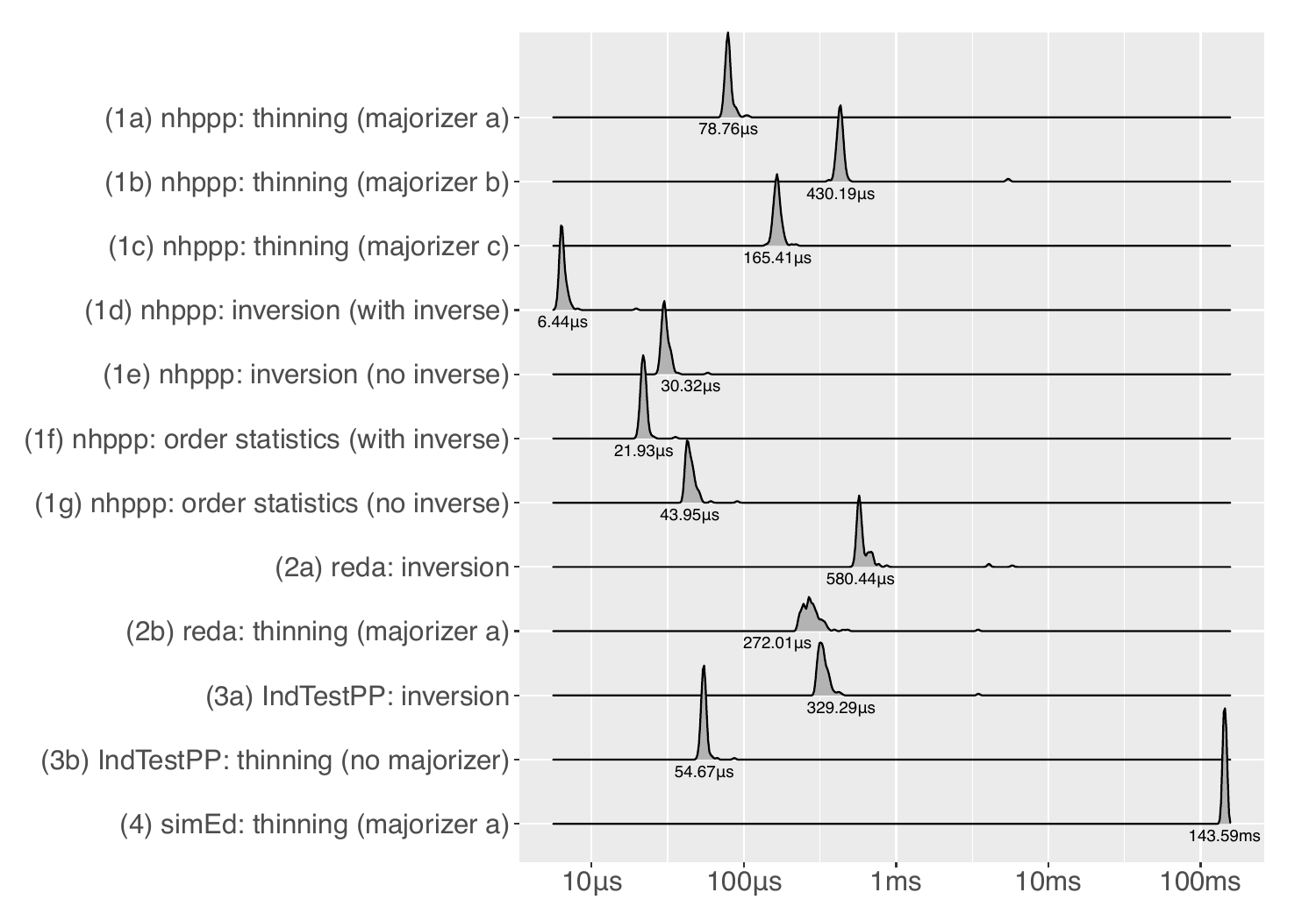} 
\end{figure}
\end{Schunk}

We provided functions with the arguments they need to run fastest. For example, functions that use the inversion or order statistics algorithm execute faster when the inverse function $\Lambda^{-1}(z)$ is provided, rather than numerically calculated, as shown in both Figures for the \pkg{nhppp} package. (Functions in other packages do not take $\Lambda(t)$ and $\Lambda^{-1}(z)$ arguments.) The fastest functions are \pkg{nhppp} functions that rely on the inversion or order statistics algorithms given $\Lambda^{-1}(z)$.

According to~\eqref{eq:thinning-efficiency}, the thinning algorithm has higher efficiency, and is expected to execute faster, for majorizer functions that envelop the intensity function more closely. Observe that $\lambda_{*a} \succ \lambda_{*c}$ and $\lambda_{*b} \succ \lambda_{*c}$ in Fig~\ref{fig:example-function-plot}. As expected, the execution times are indeed shorter for majorizer `c' compared to `b' in Figures~\ref{fig:comptimes_all_samples} and~\ref{fig:comptimes_one_sample}. However, the execution times are longer with majorizer `c' compared to `a' because \fct{draw\_intensity}, the function that uses constant majorizers, and \fct{draw\_intensity\_step}, the function that use piecewise constant majorizers, are implemented differently. \fct{draw\_intensity} happens to be faster in this example, but this is not always true.

In \pkg{nhppp}, functions that use the inversion or order statistics algorithms can exit earlier when only the first event is requested. This is not possible, however, for the thinning algorithm. This efficiency does not appear to be implemented in the other packages.

\subsubsection{Time performance of vectorized functions}\label{sec:vectorized-functions}

In \proglang{R}, `vectorized' computation, where operations are done in columns, is faster than using \code{for} loops or \fct{apply} functions. As shown in Table~\ref{tab:nhppp_functions}, \pkg{nhppp} includes vectorized functions for sampling from (i) piecewise constant intensities, using \fct{[vdraw|vztdraw]\_sc\_step\_regular}; and (ii) general intensity functions, using \fct{[vdraw|vztdraw]\_intensity\_step\_regular}. 

We compared the execution speed of non-vectorized and vectorized functions for sampling $\ensuremath{10^{5}}$ times from the piecewise constant `b' majorizer ($\lambda_{*b}$) in Fig~\ref{fig:example-function-plot}. The expected number of events with $\lambda_{*b}$ in $(0, 6\pi]$ is
$741.97$.
\red{When drawing only the earliest event, the vectorized function is approximately
$113$ times faster than the non-vectorized function (median $59 ms$ versus $6717 ms$ over $\ensuremath{10^{5}}$ simulations).
When drawing all events , the vectorized function is approximately $1.4$ times faster than the non-vectorized function (median $36.55 s$ versus $50.97 s$ over $\ensuremath{10^{5}}$ simulations). The reason that the difference in speed attenuates is that the current implementation of the vectorized functions does not use sparse matrices to store samples, which introduces inefficiencies the  expected number of samples becomes larger.}

%% -- Summary/conclusions/discussion -------------------------------------------

\section{Summary and next developments} \label{sec:summary}

The \pkg{nhppp} facilitates the simulation of NHPPPs from time-varying intensity or cumulative intensity functions. Its claim is that it (i) simulates correctly from a target density, not just from an approximation; (ii) samples conditional on observing at least one event in an interval; (iii) accomodates user provided random number stream objects; and (iv) is fast. The current version includes one vectorized function for sampling from regular-spaced piecewise constant intensity functions. In future releases we will further optimize execution speed and memory usage.

%% -- Optional special unnumbered sections -------------------------------------

\section*{Computational details and credits}

\proglang{R}~4.3.1~\citep{R-program}
was used for all analyses.
Packages
\pkg{xtable}~{1.8.4}~\citep{xtable-package} and
\pkg{knitr}~{1.45}~\citep{knitr-package}
were used for automatic report generation. Packages
\pkg{ggplot2}~{3.4.4}~\citep{ggplot2-package},
\pkg{ggridges}~{0.5.5}~\citep{ggridges-package}, and
\pkg{latex2exp}~{0.9.6}~\citep{latex2exp-package}
were used for plot generation and \LaTeX formatting.
Packages
\pkg{nhppp}~{0.1.4}~\citep{nhppp-package},
\pkg{bench}~{1.1.3}~\citep{bench-package},
\pkg{rstream}~{1.3.7}~\citep{rstream-package},
\pkg{otinference}~{0.1.0}~\citep{otinference-package}, and
\pkg{parallel}~{4.3.1}
were used in the examples and the analyses.

All computations were done on an Apple M1 Max machine with 64 megabytes of random access memory. 
%A preprint of the current paper is in~\cite{trikalinos2024nhppp}.
%
\proglang{R} itself
and all aforementioned packages are available from the Comprehensive
\proglang{R} Archive Network (CRAN) at
\url{https://CRAN.R-project.org/}.

\section*{Acknowledgments}
This work was funded from grant U01CA265750 from the National Cancer Institute.
We thank the investigators of the Cancer Incidence and Surveillance Modeling Network (CISNET)
Bladder Cancer Site Stavroula Chrysanthopoulou, Jonah Popp, Fernando Alarid-Escudero,
Hawre Jalal, and David Garibay for useful discussions.

%% -- Bibliography -------------------------------------------------------------
%% - References need to be provided in a .bib BibTeX database.
%% - All references should be made with \cite, \citet, \citep, \citealp etc.
%%   (and never hard-coded). See the FAQ for details.
%% - JSS-specific markup (\proglang, \pkg, \code) should be used in the .bib.
%% - Titles in the .bib should be in title case.
%% - DOIs should be included where available.

\newpage
\bibliography{refs}

%% -- Appendix (if any) --------------------------------------------------------
%% - After the bibliography with page break.
%% - With proper section titles and _not_ just "Appendix".

\newpage
\begin{appendix}

\section{Piecewise constant majorizer functions}\label{app:piecewise_majorizer}

Let $\lambda(t)$ be either a monotonic (and possibly non-continuous) function, or if it is non-monotonic, a $K$-Lipschitz continuous intensity function, i.e., an intensity function where $|(\lambda(b) - \lambda(a))| \le K|b-a|$, with $K$ known.   Algorithm~\ref{alg:lambda_majorizer} finds a piecewise constant majorizing function $\lambda_*(t)$. Starting from a partition of the time interval in time steps (not necessarily equal) it finds an upper bound for $\lambda$ within the each partition.

If $\lambda(t)$ is monotonic, the least upper bound (supremum) is always found at the extremes of the interval and no knowledge of $K$ is required.

The algorithm should be started with a good partitioning of the time interval. In practice, it is generally easy to specify equispaced intervals that are fine enough and impose little computational penalty for the application.

The \pkg{nhppp}
function \fct{get\_step\_majorizer} implements Algorithm~\ref{alg:lambda_majorizer}. %Functions \fct{draw\_intensity\_step}, \fct{draw\_sc\_step}, \fct{draw\_sc\_step\_regular} and \fct{vdraw\_sc\_step\_regular} expect the majorizer function values as an argument.

\begin{Schunk}
\begin{Sinput}
R> get_step_majorizer(
+    fun = abs, breaks = -5:5, is_monotone = FALSE,
+    K = 1
+  )
\end{Sinput}
\begin{Soutput}
 [1] 5.5 4.5 3.5 2.5 1.5 1.5 2.5 3.5 4.5 5.5
\end{Soutput}
\end{Schunk}

 \begin{algorithm}[h!]
\caption{Pick a majorizing piecewise constant function $\lambda_*(t)$. Partition the interval and find an upper bound for $\lambda(t)$ in each partition.}\label{alg:lambda_majorizer}
\begin{algorithmic}[1]
\Require{
\begin{itemize}
\item[] { $ $ }
\item[] { $\lambda(t)$ is $K$-Lipschitz in $(a, b]$ } 
\item[] { Partition interval: $(a, b] = \bigcup_{m=1}^{M} (a_m, b_m] $  \Comment{$a=a_1, b_M = b, a_{m} = b_{m-1} \ (m>1)$}} 
\end{itemize}}
\State $c \gets K$ \Comment{Fastest possible slope}
\If{ $\lambda(t)$ is monotonic}  \Comment{Then $\sup_{t\in (a_m, b_m]} (\lambda(t))= \max\big(\lambda(a_m), \lambda(b_m)\big)$}  
\State $c \gets 0$    
\EndIf
\For{$m \in [M]$}:
    \State $\lambda^*_m \gets \max\big(\lambda(a_m), \lambda(b_m)\big) + c(b_m - a_m)/2$ \Comment{Upper bound for $\lambda(t)$ in $(a_m, b_m]$}
\EndFor
\State{$\lambda_*(t) \gets \bigcup_{m=1}^{M} \{ \big( (a_m, b_m], \lambda_m \big) \} $} \Comment{Piecewise constant map: $\lambda: (a_m, b_m] \mapsto \lambda_m$}
\State
\Return{ $\lambda_*(t)$  }
\end{algorithmic}
\end{algorithm}

\newpage
\section{Conditional sampling from NHPPPs} \label{app:conditional_sampling}
Algorithm~\ref{alg:NHPPP_conditional} is a direct modification of the order statistics Algorithm~\ref{alg:NHPPP_order_stats} to sample conditional on observing $m$ events in $(a,b]$. (The modification is in line 1 in red font.)
To sample exactly $m$ points, change line 1 of Algorithm~\ref{alg:NHPPP_conditional} to
\begin{center}
$N \gets m$.
\end{center}
To sample up to $k$ earliest points, replace line 11 with in Algorithm~\ref{alg:NHPPP_conditional} with
\begin{center}
\textbf{return} {$\{Z_{(i)} \ | \ i \le k, Z_{(i)} \in \mathcal{Z}\}$}.
\end{center}

\begin{algorithm}[h!]
\caption{Modified order statistics algorithm for sampling at least $m$ events from an NHPPP given $\Lambda(t), \Lambda^{-1}(z)$.}\label{alg:NHPPP_conditional}
\begin{algorithmic}[1]
\Require $\Lambda(t), \Lambda^{-1}(z), t \in (a, b]$ \Comment{$\Lambda^{-1}(z)$ possibly numerically}
\State \textcolor{red}{$N \gets N \sim \textrm{TruncatedPoisson}_{N \ge m}\big(\Lambda(b)-\Lambda(a)\big)$ \Comment{$(m-1)$-truncated Poisson}}
\State $t \gets a$
\State $\mathcal{Z} \gets \emptyset$ \Comment{$\mathcal{Z}$ is an ordered set}
\If{N > 0}
    \For{$i \in [N]$}:
        \State $U_i \gets U_i \sim \textrm{Uniform(0, 1)}$ \Comment{Generate order statistics}
        \State $\mathcal{Z} \gets \mathcal{Z} \cup \{ \Lambda^{-1} \Big( \Lambda(a) + U_i \big( \Lambda(b) - \Lambda(a) \big)\Big) \} $ 
    \EndFor
    \State $\mathcal{Z} \gets \textrm{sort}(\mathcal{Z})$ 
\EndIf
\State
\Return{$\mathcal{Z}$} \Comment{Up to $k$ earliest points: \textbf{return} $\{Z_{(i)} \ | \ i \le k\ , Z_{(i)} \in \mathcal{Z} \}$}
\end{algorithmic}
\end{algorithm}

\end{appendix}
\newpage
%% -----------------------------------------------------------------------------

\end{document}